\documentclass[letterpaper,11pt]{article}
\usepackage{jheppub}
\usepackage{amsmath,latexsym}
\usepackage{graphicx}      
\usepackage{float}
\usepackage{tikz-feynman}  
\tikzfeynmanset{compat=1.1.0,warn luatex=false}  
\usepackage{booktabs}
\usepackage{comment}
\usepackage{setspace}
\usepackage{amssymb}
\usepackage{amstext}
\usepackage{amsfonts}
\usepackage{bbold}
\usepackage{slashed}
\usepackage{hyperref}
\usepackage{cleveref}
\usepackage{xcolor}
\usepackage{cancel}
\usepackage{physics}
\usepackage[normalem]{ulem}
 \usepackage{tikz}
 \usetikzlibrary{decorations.pathmorphing,arrows.meta}
\newcommand{\beq}{\begin{equation}}
\newcommand{\eeq}{\end{equation}}

\newcommand{\bea}{\begin{eqnarray}}
\newcommand{\eea}{\end{eqnarray}}

\newcommand{\nn}{\nonumber}

\newcommand{\be}{\begin{eqnarray}}
\newcommand{\ee}{\end{eqnarray}}

\definecolor{SGreen}{rgb}{0.0547,0.613,0.328}
\definecolor{NodeBlue}{rgb}{0.0547,0.148,0.578}

\begin{document}
\emergencystretch=3em
\renewcommand{\theHfigure}{\thesection.\arabic{figure}}

\title{A Unified Treatment of the  Self-Force Problem} 

\author[a]{Beka Modrekiladze,}
\author[b]{Ira Z. Rothstein,}
\author[b]{and Jordan Wilson-Gerow}

\affiliation[a]{Deutsches Elektronen-Synchrotron DESY, Notkestra{\ss}e 85, 22607 Hamburg, Germany.}
\affiliation[b]{Department of Physics, Carnegie Mellon University, Pittsburgh, PA 15213, USA}

\abstract{
This paper introduces formalism, based on worldline effective field theory, which allows one to systematically calculate the gravitational self-force
on a compact object immersed in an environment  with non-vanishing stress energy. Using the closed time path integral we present  a universal effective action that can
be used to calculate the  equations of motion of a compact object due to its interaction with the environment
to any order in the relevant expansion parameters, the relative importance of which depends upon the choice of environment. We show that the leading order equations of
motion can be universally written in terms of 
 the retarded two-point function of the stress-energy tensor for 
the  environment. 
The resulting action can be used to calculate both dissipative (dynamical friction) and conservative forces in a completely relativistic fashion.
We demonstrate the utility of the result by calculating the dissipative force due to dust, an inviscid fluid, and a coherent field. 
Our results agree with those previously derived in the literature. We furthermore show that the famous ``Coulomb Log" found by Chandrasekhar should be interpreted as a renormalization group log due to a UV divergence that renormalizes the dissipative part of the in-in action.
We then prove that this log is universal in the Newtonian limit in that its value is universal,  for a generic class of environments and trajectories. This log is the leading log in an RG
flow in the dissipative action that has yet to be explored.

}

\maketitle

\section{Introduction}\label{sec: 1}

The self-force problem can be loosely defined as follows.
We consider a localized object $O$  interacting with some external system $E$, which may or may not break Poincar\'{e} invariance. 
For instance $E$ may be a field which couples to the object, such as the metric or the photon, or it could
be some source  with non-vanishing stress-energy. 
The motion of $O$ perturbs $E$ which then reacts back on $O$. The force may act locally/nonlocally in time, i.e., the force is Markovian/non-Markovian.
The latter case implies the force will depend upon the past trajectory of the particle.
The canonical example of a local force is a radiation
reaction force in a linearized theory such as  the ALD \cite{Abraham1905,Dirac1938,Lorentz1892} force in electrodynamics
\beq
\label{ALD}
m a^\mu
=
F^\mu_{\text{ext}}
+
\frac{2 e^2}{3}
\left(
\dot{a}^\mu
+
a^\nu a_\nu \, u^\mu
\right)
\eeq
or the Burke--Thorne force in GR \cite{BurkeThorne1970}, both of which act locally in time.

 The term ``self-force" has typically been applied in the context of general relativity in situations where $E$ is a curved vacuum spacetime~\cite{Poisson:2011nh,Barack:2018yvs,Poisson:2003nc}, however we will use the term more broadly. 
While gravitational self-force has previously been formulated for non-vacuum spacetimes to include additional massless fields~\cite{Zimmerman_2014,Linz:2014vja, Zimmerman_2015}, and to describe orbits outside of fluid stars~\cite{Isoyama:2015zbe}, in this work we address the case where $E$ contains matter fields at finite density and $O$ is embedded within $E$, so that the environment breaks local Lorentz invariance. Our formalism unifies previously considered cases and allows for the simultaneous treatment
of vacuum and non-vacuum effects.

When Lorentz invariance is unbroken by the environment,  nonlocal-in-time forces can only arise due to non-linearities, at least for massless fields in three spatial dimensions, 
as the retarded propagator support is isolated on the light-cone.\footnote{In even spatial dimensions the retarded propagator is proportional to $\theta(x^2)$ as opposed to
$\delta(x^2)$ in 3+1.} However, for Poincar\'{e} violating environments nonlocalities will arise even 
in the linearized theory. Physically, this is due to the  fact  that the propagation speed of the disturbance is fixed by the bulk properties of the system. For instance, for a fluid the  object speed can be larger than the sound speed and the particle will be perturbed by a disturbance it generated
at an earlier time.

The classic example of this is ``dynamical friction", 
defined as the drag on a compact object moving through a gravitating  medium due to energy lost to the environment. The effect is interpreted as being due to the formation of a wake (overdensity) behind the object.\footnote{This is opposed to a pressure under-density in the wake of a boat.}
Dynamical friction plays a central role in astrophysics, driving the sinking of satellites and the assembly of supermassive black hole binaries~\cite{Volonteri:2010wz}. In the gravitational wave era it has also become a precision observable: over the many orbital cycles accumulated in the bands of Einstein Telescope and LISA~\cite{Punturo:2010zz,LISA:2017pwj}, the drag from an accretion disk~\cite{Yunes:2011ws,Speri:2022upm}, a dark matter spike~\cite{Gondolo:1999ef,Merritt:2002vj,Macedo:2013qea}, or an ultralight scalar cloud~\cite{Barsanti:2021ydd} leads to a measurable dephasing of the waveform. Since in matched filtering  the phase is a relevant observable~\cite{Maggiore:2007ulw}, an unmodeled drag biases source parameters and can even mimic a violation of general relativity~\cite{Cardoso:2019rou,Gupta:2024gun}. Precision predictions for dynamical friction have become increasingly relevant for gravitational wave astronomy.

The original calculation of dynamical friction, due to Chandrasekhar, was performed in the Newtonian limit, and assumed a pressureless dust for the background with a straight-line trajectory for the object. It was improved via the inclusion of post-Newtonian corrections in~\cite{Lee:1969ApJ155687} and generalized to the relativistic case in~\cite{Syer:1994MNRAS270205}.
The problem of a fluid (hydrodynamic) background with a straight-line trajectory was first considered in~\cite{Ruderman:1971GalacticWakes} in the Newtonian limit, followed by ~\cite{Rephaeli:1980FlowPast, Petrich:1989} in the supersonic and relativistic cases, and later revisited in~\cite{Ostriker1999} for the
subsonic case. It was subsequently applied to circular orbits in~\cite{KimKim2007}, which was soon after lifted to relativistic circular motion in~\cite{Barausse:2007ph}, and generalized to elliptical orbits in~\cite{buehler2023ecc}. More recently, the interest in environmental effects for extreme-mass-ratio inspirals has led to formulations of the dynamical friction problem for a secondary body in motion around a primary black hole~\cite{Brito:2023pyl, Dyson:2025dlj, Datta:2025ruh}. 

We pause here to note that historically all of these analyses  assume 
the point particle approximation with no mention (as far as the authors are aware)
of the finite-size effects, which grow with the relevant gradients.\footnote{Lorentz contraction
enhances finite-size (gradient) effects.} A complement to these approaches has been the recent fully general relativistic computations for black holes in uniform motion through homogeneous clouds of collisionless particles and coherent scalar fields~\cite{Boudon:2023supersonic,Vicente:2022ivh, Traykova:2023qyv, Dyson:2024qrq}. These calculations automatically include finite-size effects, such as accretion, however, they are limited in their applicability as they apply only to straight-line motion.

As we shall discuss, our formalism, being an outgrowth of the previous formalism
\cite{GoldbergerRothstein2006,Goldberger:2006bd,Goldberger:2005bh,Goldberger:2020horizon} will allow for systematically including such finite-size corrections. 
Our method of calculation differs from previous work in multiple ways, the most significant of which is that it is based upon an EFT which is systematically improvable. An EFT approach is essential for cleanly explaining how to treat the UV divergences in the previous computations in the literature. Moreover, the EFT approach provides a framework to incorporate the exact GR computations~\cite{Vicente:2022ivh, Traykova:2023qyv, Dyson:2024qrq} into a framework where they can be extended beyond straight-line motion. 

The EFT enables one to extract the relevant short distance parameters using a simple configuration, which can then be used in more complicated situations. For instance, in vacuum one uses precise computations in General Relativity to determine near-zone properties such as the tidal response of a black hole~\cite{Binnington:2009bb, Damour:2009vw}, which determines Wilson coefficients in the point-particle EFT~\cite{Kol:2011vg} (Love numbers).
Once these coefficients have been fixed they can be used  to
calculate the induced potentials due to tidal deformations.
We can do something similar for dynamical friction, the exact GR computations of~\cite{Vicente:2022ivh, Traykova:2023qyv, Dyson:2024qrq} for straight-line motion through dust, can be used to extract the ``near-zone/UV'' data, which fixes Wilson coefficients in the EFT, as will be done in section 6.4, which in turn can be used to calculate the force for some generalized trajectory. 

 Previous work on the subject has used different methods depending on the medium. 
For collisionless dust, one can sum over the cumulative scattering events of the background particles \cite{Chandrasekhar1943}. 
For a fluid, one instead solves the linearized Euler and gravitational equations, whose solution exhibits a wake that sources the drag force \cite{Ostriker1999,KimKim2007,Petrich:1989,Barausse:2007ph}. Our approach is closer in spirit to vacuum self-force methods: we integrate out the environment and compute an effective action. 
At linearized order this yields a universal expression for the force in terms of the medium stress-tensor response function. 
In this language, the wake geometry need not be analyzed separately; its effects are encoded in the effective action, including the time nonlocality associated with the finite propagation speed. 
Moreover, there is no obstruction to extending the calculation beyond linearity, provided one has the appropriate EFT description of the bulk.

Thinking in terms of the response function clarifies the nature of the force law as it allows one to simply distinguish between the conservative and dissipative as well as local and nonlocal 
contributions.\footnote{The term  ``dynamical friction'' refers to the dissipative force.}  Both conservative and dissipative forces
may be either local or nonlocal.
 Considering a generic linear response the force will be given by
\beq
F^i(t,x)=  \int d^3x^\prime dt^\prime \chi^{ij}(t-t^\prime,x-x^\prime) J^j(t^\prime,x^\prime),
\eeq
where the source is the velocity field \footnote{Since we are interested in the gravitational response and the stress-tensor's spatial components will
be proportional to the velocity.}
\beq 
J^i=u^i(t,x)=\int d\lambda \,\dot x^i(\lambda)\delta^4(x-x(\lambda)).
\eeq

Whether or not this leads to a local or nonlocal force  depends upon the nature of the response function. 
In particular, whether or not its expansion in powers of the time difference is convergent. This is most easily seen in coordinate space.
Consider, for instance, the case of flat 3+1 dimensions with a vacuum environment: by causality the response at some affine time $z$ will be weighted by $\int dz^\prime \delta((x(z)-x(z^\prime))^2)$, and
choosing to parameterize the worldline by $w=(x(z)-x(z^\prime))^2$, we find  that the force is given by
\beq
f \sim \int dw\, G(w) \frac{\delta(w)}{w}.
\eeq
$G(w)$ is necessarily an analytic function in $w$ since it is fixed by the interaction vertices of the local action. Since timelike worldlines can only intersect the light-cone of their emitted radiation at one point, we can expand about the coincidence limit. This leads
to a finite series as anything beyond linear order in $w$ will vanish. That is why the ALD force (\ref{ALD}), is exact, up to finite-size effects (finite-size effects were discussed in \cite{Galley:2010xn,Forgacs:2012uq,Galley:2012vn}).
No small derivative approximation is made.
We can also see from this analysis that there can be no log divergences, i.e. no non-trivial renormalization.
However, as we shall see, when the vacuum breaks spacetime symmetries,  novel renormalization group phenomena can arise.

When Lorentz invariance is broken by the environment, the propagator can have support away from the vacuum light cone. For instance, in an ideal fluid, the sound pole replaces the light-cone singularity $\delta(x^2)$ with an acoustic-cone singularity proportional to $\delta(c_s^2t^2-r^2)$. For supersonic motion, $\dot{x}>c_s$, the worldline can then non-trivially intersect its previously generated acoustic cone. Disturbances in more general media can also have support throughout the interior of the acoustic cone. As a consequence there can be history-dependent (hereditary) contributions to the self-force, even in flat spacetime.

At a technical level, the strict $\delta(w)$ that truncated the Taylor expansion of $G(w)$ to finitely many terms is no longer present. As a result, the proper-time integral no longer terminates, the momentum integral is no longer polynomial in $k$, and logarithmic divergences $\int dk/k$ become possible. These logarithms are the avatar of RG running and necessitate the introduction of counterterms in the worldline action. In particular, as we shall see,  a conservative action  will generate a counter-term for a dissipative (time reversal violating) operator.
This possibility arises as a consequence of the fact that the system is properly described by the closed-time path integral
that treats time in an asymmetric fashion. As a result, the one-loop correction from the conservative (real) part of the effective action generates a logarithmic divergence that must be absorbed by the dissipative counter-term $K(\mu)$, inducing RG flow even when the tree-level dissipative coupling vanishes.
 
 It is useful to decompose
the linear response into real and imaginary parts in
Fourier space 
\beq
\chi_{ij}(\omega,k)= \chi_{ij}^R(\omega,k)+i \chi_{ij}^I(\omega,k).
\eeq
Assuming there is no explicit breaking of time reversal,
the real and imaginary parts will conserve and violate this symmetry, respectively.
The power loss (again in the linearized approximation) is given by
\beq
P \sim  \omega J_i(\omega) \chi^I_{ij}(\omega) J_j(-\omega).
\eeq

Given that the generalized force is nonlocal in time the result will depend upon the time interval over which one integrates, as well as the nature of the trajectory. The original calculations were done for straight-line trajectories which are not self-consistent in the sense that the trajectory is assumed and the force is then calculated based upon that fixed trajectory.
For infinite time 
there is no loss when the motion is subsonic, as there is no Cherenkov radiation. 
However, for a finite time interval, energy will be lost
to building a wake. Strictly speaking this force is not entirely dissipative but on short time
scales this cannot be distinguished.
This is as opposed to the supersonic where the system unambiguously loses energy due to radiation lost at infinity. 

\subsection{Overview}

In this work we use an EFT approach to the generalized self-force problem which
gives a universal formula for
the effective action (and force) parameterized by the two-point function of the stress energy tensor in medium. This result will be valid at leading order, and there is no obstruction to using the framework for calculating beyond leading order, where universality is lost. 

In addition to the universal results we present, a primary novelty of this work is the use of the generalized point-particle action of \cite{Modrekiladze:2024htc}. To have a controlled perturbative description of dynamical friction the inclusion of ``finite-size effects'' described by this action is non-negotiable; there is an ultraviolet divergence already present in the leading order Chandrasekhar result whose systematic understanding necessitates  renormalization and EFT matching. Historically the treatment of this divergence has been ad hoc, but by employing the EFT introduced in~\cite{Modrekiladze:2024htc} one can proceed systematically.

The problem is amenable to a perturbative treatment via EFT thanks to a hierarchy of physical scales, and we detail this power counting in \cref{sec:powercounting}. In \cref{WLEFT} we cover the essential details of the point-particle EFT. Section \ref{sec:universalLOresult} covers the derivation of the 1-loop integral computing  dynamical friction for a general environment and trajectory, and \cref{sec:UVandRG} proves the universality of the logarithmic divergence in this integral. In \cref{sec:fluidexample} we provide an example application of these results to a fluid environment, recovering established results and placing a novel emphasis on renormalization.

We work in units where $c=\hbar=1$ and  adopt the mostly minus metric convention. Both Latin and Greek indices range from $0-3$. We use the compact momentum integral notation, $\int_{k}\equiv\frac{d^{4}k}{(2\pi)^{4}}$.

\section{Power Counting}\label{sec:powercounting}
Let us first consider the relative importance of environmental versus vacuum (radiation-reaction) effects\footnote{For a review of environmental effects see \cite{Barausse:2015EnvironmentalReview}}. For a binary system, the Burke-Thorne force for orbital motion is a 2.5PN effect,
\beq
F_{BT} \sim v^5\frac{Gm^2}{r^2}.
\eeq
whereas the leading order environmental effect, treated relativistically, is \cite{Petrich:1989}, 
\beq
F^{>}_{\rm env} \sim 4\pi G^2 m^2 (p_E+\rho_E)\,\frac{\gamma^2(1+v^2)^2}{v^2}\,\ln\!\left(\frac{b_{\rm max}}{b_{\rm min}}\right),
\eeq
for supersonic motion.
Thus the vacuum radiation-reaction force is suppressed by additional powers of the velocity relative to the leading environmental force \footnote{At least when $\rho_E$ is sufficiently large.}. The factor $v^5$ simply reflects the post-Newtonian suppression of the Burke-Thorne force. In practice, this means that environmental effects are expected to be most relevant during the earlier stages of an inspiral, while vacuum radiation reaction becomes increasingly important closer to merger. Which regime is more phenomenologically relevant depends on the detector band and on the properties of the environment.

It is also useful to compare dynamical friction with mass accretion. Using the Bondi--Hoyle--Lyttleton estimate for a black hole \cite{HoyleLyttleton1939,BondiHoyle1944,Edgar:2004BHLReview}, one has
\beq
\dot m_{\rm BHL} \sim 4\pi \rho_E \frac{(Gm)^2}{(v^2+c_s^2)^{3/2}},
\eeq
so the corresponding drag due to mass accretion scales as
\beq
F_{\rm acc} \sim \dot m_{\rm BHL} \, v \sim 4\pi \rho_E (Gm)^2\,\frac{v}{(v^2+c_s^2)^{3/2}}.
\eeq
Comparing this with the dynamical friction force, we see that in the supersonic regime the two effects can be of similar parametric size, differing mainly by the Coulomb logarithm and order-one coefficients, whereas in the subsonic regime the relative importance depends sensitively on the sound speed and on the detailed structure of the medium. This is consistent with the standard astrophysical discussion of the interplay between gaseous dynamical friction and accretion drag \cite{LeeStahler2011}. A systematic EFT approach to accretion will be treated in a separate paper \cite{GMRII}.

As far as the approximations made in this paper are concerned, even in vacuum,
we demand finite-size effects to be small. The first expansion parameter is
therefore $R/\alpha$, where $R$ is the size of the compact object and $\alpha$
denotes a generic IR length scale. In a medium, the relevant IR scale can be set
by gradients of the background density, for example $\alpha\sim \rho/|\partial \rho|$.
When the object moves relativistically with respect to the medium Lorentz
contraction enhances finite-size effects.  A disturbance with rest-frame
wavelength of order $\alpha$ is shortened to $\alpha/\gamma$ in the object frame,
so the effective expansion parameter is
\beq
\label{e1}
\epsilon_0\equiv  \frac{\gamma R}{\alpha}.
\eeq

We will be expanding around flat space, and as such, all curvature corrections either due to the object or the environment should be small, which therefore
bounds
\beq
\label{e2}
\epsilon_1\equiv Gm\gamma/\alpha\ll 1\,.
\eeq
For a black hole or neutron star this parameter is redundant with $\epsilon_0$. The environmental mass density is bounded similarly
\beq
\label{e3}
\epsilon_2\equiv  G \rho_E L^2 \ll 1,
\eeq
where $L$ is the range of support of the environment. These parameters parallel those considered for extreme mass-ratio systems in~\cite{Brito:2023pyl, Dyson:2025dlj, Datta:2025ruh}.

For a fixed density there is an IR cutoff since
the size of the environment is bounded by the Jeans length as a stability criterion.
For a fluid
\beq
\lambda_J=\frac{c_s}{\sqrt{4\pi G_N(\rho_E+p_E)}},
\eeq
Since $c_s\ll 1$, this is a stronger bound on $L$ than the 
flatness constraint.
For a collisionless dust, a velocity dispersion $\sigma$ plays the role of 
 pressure which stabilizes the system giving 
an effective Jeans length
\begin{equation}
\lambda_J^{\mathrm{eff}}= \frac{\sigma}{\sqrt{4\pi G_N\rho_E}}.
\end{equation}
The cloud will not collapse if the crossing time ($L/\sigma$) is short compared
to the collapse time $t_{\mathrm{col}}= \sqrt{\frac{1}{\rho_E G_N}}$.
For a coherent scalar field of mass $m_s$, the pressure is supplied by gradient energy. The Jeans wavenumber for the Schr\"odinger--Poisson system is \cite{Hu:2000ke,Hui:2016ltb}
\beq
k_J = (16\pi G_N \rho_E\, m_s^2)^{1/4},
\eeq
giving a Jeans length $\lambda_J^{\rm scalar} = 2\pi/k_J$ that bounds the maximum size of a stable homogeneous condensate and provides the IR cutoff for the scalar field dynamical friction calculation.

\section{The Worldline Effective Field Theory in a Medium}
\label{WLEFT}
The formalism generated here is the natural sequel to  \cite{GoldbergerRothstein2006,Goldberger:2006bd}, where it was shown that by working in an EFT, treating one scale at a time, one can simplify the problem of
radiating systems of compact objects.  The EFT is based upon an expansion in $\omega R$ where $R$ is the typical radius of the bodies, and $\omega$ is the frequency of the radiation.
In the non-relativistic, small curvature limit, one decomposes the graviton into   potential and radiation modes which leads to a factorization that allows for the systematic resummation
of logarithmic terms via renormalization group techniques. The finite-size effects are accounted for by the inclusion of higher-dimensional operators which also act
as counterterms for UV divergences which inevitably arise when working in the point particle limit.  

The action for the worldline EFT in vacuum is given by
\beq
S= -m \int d\lambda \sqrt{\dot x^2} + \alpha_E \int E^2 d\lambda + \alpha_B \int B^2 d\lambda + \ldots
\eeq 
Here $E^2\equiv E_{\mu\nu}E^{\mu\nu}$ and $B^2\equiv B_{\mu\nu}B^{\mu\nu}$, where the electric and magnetic components of the Weyl tensor are
\beq
E_{\mu\nu}=C_{\mu\alpha\nu\beta}v^\alpha v^\beta,
\qquad
B_{\mu\nu}=\frac{1}{2}\epsilon_{\mu\alpha\rho\sigma}C^{\rho\sigma}{}_{\nu\beta}v^\alpha v^\beta .
\eeq
$\alpha_{E,B}$ are the leading-order finite-size corrections, and define a gauge-invariant notion of Love numbers.\footnote{Since $B_{\mu\nu}$ is parity
odd, an object whose internal structure violates parity admits the additional
operator $\alpha_{EB}\int E_{\mu\nu}B^{\mu\nu}\,d\lambda$, defining a
parity-violating analogue of the Love numbers. The consequences of such
couplings were studied in~\cite{Modrekiladze:2022ioh}.}
Higher order terms will be down by powers of the radius of the object
which is taken to be small compared to the scales which it probes.
 In this paper we will ignore the finite-size effects
on the drag forces, which have yet to be calculated as far as we are aware.

\subsection{The Conservative Action}

As discussed in \cite{Modrekiladze:2024htc}, the action for a compact object, to leading order
in an expansion in $\omega R$, is constrained by the symmetries. Before specifying a particular medium, the most general conservative action consistent with worldline
reparameterization invariance and general coordinate invariance allows the local worldline mass to depend on the scalar data supplied by the environment:
\beq
\label{conserve}
S_{\text{kin}}
= - \int d\lambda\,
\sqrt{\dot x^2}\,M\!\left(\dot x;\Psi,\mu\right) .
\eeq
Here $\Psi$ denotes the collective background fields of the medium, and $M$ is a function of all scalars that can be built from $\Psi$ and $\dot x^\mu$. For example, once the medium is specialized to a fluid with four-velocity $u^\mu$ and density $\rho$, this set can include
\beq
\gamma = \frac{u \cdot \dot x}{\sqrt{\dot x^2}},
\eeq
as well as $\rho$ and derivative corrections. Any dependence of $M$ on velocity-dependent invariants such as $\gamma$ changes the
mass shell constraint since, in general,
$p^\mu \neq M \frac{\dot x^\mu}{\sqrt{\dot x^2}}$.
Nonetheless the existence of RPI ensures that this
 constraint will still generate
gauge reparameterizations.

Since we will eventually allow for dissipation we will re-write this action
in the doubled closed-time-path (CTP/in-in) formalism where
the path and the degrees of freedom are doubled such that we can write the world-line action as
\beq
S= \int d\lambda_+L(\dot x_+) - \int d\lambda_-L(\dot x_-)
\eeq
There are 
two worldline reparameterization invariances corresponding to the forward and backward paths.
In flat spacetime it is convenient to work in the Keldysh basis
\beq
x_r=\frac{1}{2}(x_++x_-)~~~~~x_a= x_+-x_-,
\eeq
then the  equations of motion follow by varying the action with respect to $x_a$ and then setting it to zero.
Thus we need only expand the action to linear order in $x_a^\mu$, so we may write, to leading
order in gradients of the medium,
\beq
\label{eq:consWLaction}
S_{\text{kin}}
= - \int d\lambda\,
\dot x_a^\mu \frac{\partial}{\partial \dot x_r^\mu}
\left[\sqrt{\dot x_r^2}\,
M\!\left(\dot x_r;\Psi_r,\mu\right)\right]+O(\dot x_a^2).
\eeq
\subsection{The Dissipative Action}
What is interesting about the CTP EFT is that we are not done at this point, as we must write down all terms in the action consistent with the symmetries and the power counting.
Thus while Poincar\'{e} invariance normally does not allow for explicit dependence on $x_\pm$, $x_a$ dependence will be allowed, once it is properly defined in curved space.
In fact, if we want a dissipative action we need to
be able to write it in terms
of $x_a$.
The challenge is then to tensorialize $x_a$. 
This problem was addressed in \cite{Modrekiladze:2024htc}, which is briefly reviewed in \cref{app:diff-invariant-in-in}, where it is shown that, at least as far as the classical equations of motion are concerned, we may treat $x_a$ as a tensor. However, RPI invariance requires that we remove the component of $x_a^\mu$ parallel to the worldline tangent vector. The reason is that a separation of the two CTP histories along $\dot x_r^\mu$ can be generated by an off-diagonal reparameterization, and is therefore not an independent physical displacement. The physical Keldysh displacement is the transverse part, obtained by projecting $x_a^\mu$ onto the subspace orthogonal to $\dot x_r^\mu$ via \footnote{For the usual time path  the force is only transverse when the parameter is affine, here we are forced to project to make the action RPI. }
$X_a^\mu= P^\mu_\nu x_a^\nu$,
where the projector is given by
\beq
\label{projector_def}
P^\mu_\nu= \left(\delta_\nu^\mu-\frac{\dot x_r^\mu  \dot x_{r\nu}}{\dot x_r^2}\right).
\eeq

Now we may build the leading order action allowing for dissipation, which is given by
\beq
\label{Sdis}
S_{\text{dis}}
= \int d\lambda \;K(\dot x_r;\Psi_r,\mu)
\frac{u_r \cdot X_a}{\sqrt{\dot{x}_r^2}}\,,
\eeq
$u_r$ will be the velocity for a fluid or dust, or $\partial_\mu \phi$ for a coherent field.
This is just the canonical friction term. For instance, for a relativistic damped oscillator $u^\mu$ picks out the rest frame of the oscillator so in its
rest frame the action would be $X_a^0$ which by its transverse nature is fixed
to be ${\vec v_r} \cdot \vec x_a$, giving  $\vec F \sim \vec v$.

\section{The Universal Leading Order Result}\label{sec:universalLOresult}

Equation~\eqref{eq:consWLaction} is the action for the point particle prior to integrating out the graviton and the environment, assuming no dissipation. We now wish to calculate the effective in-in action, obtained by integrating out both the environment and metric perturbations. This will then give us the effective equations of motion. We will explicitly compute the effective action to leading order in Newton's constant $G$, maintaining Lorentz covariance. That is, we compute the leading diagram, \cref{fig:my-diagram}, in the post-Minkowskian rather than post-Newtonian expansion.
It is important to recall from our power counting section that the theory breaks down in the ultra-relativistic limit, as can be seen from eqs.~(\ref{e1}-\ref{e3}).

\subsection{Integrating out the dynamical fields}\label{sec:integratingout}

Our system consists of a point-particle, dynamical metric, and a general matter environment. To obtain the effective action for the particle we will integrate out both the environment and the gravitational field, proceeding in two steps: first we will integrate out a general environment to obtain an effective gravity theory, and then integrate out the metric. To consistently account for conservative and dissipative effects, we will use the CTP formalism. 

 Readers familiar with the CTP formalism can skip to \cref{sec:LOdiagram}. The upshot of this section can be summarized as follows.  The leading order CTP effective action can be obtained from the standard leading order in-out effective action by simply taking the diagram that computes the in-out effective action, replacing Feynman propagators by retarded propagators, and replacing one of the point-particle energy momentum tensors by $T^{\mu\nu}_{a}$ defined below.\footnote{One must also multiply by two, to correct the symmetry factor of the diagram.}  The in-out effective action is simply the one-loop gravitational self-energy of a point particle 
 which accounts for the graviton mixing with the environment, via an  insertion of the two point function $\langle\delta T^{\mu\nu} \delta T^{\alpha\beta}\rangle$, which describes the linear response of the environment.
 This is depicted in \cref{fig:my-diagram}.

While our focus is entirely classical, it is convenient to use the language/notation of path-integrals and take the classical limit. The contributions to the worldline effective action from its gravitational interactions with the environment can be written as
\begin{align}\label{eq:SKWLIeff}
    &e^{i I_{\mathrm{eff}}[x_+,x_-]} = \nonumber \\
    &\oint\mathcal{D}g_{+}\mathcal{D}g_{-} \oint \mathcal{D}\Psi_{+} \mathcal{D}\Psi_{-}\,e^{iS_{EH}[g_+]+iS_{env}[\Psi_+,g_+]+iS_{kin}[x_+,g_+,\Psi_{+}
    ]-iS_{EH}[g_-]-iS_{env}[\Psi_-,g_-]-iS_{kin}[x_-,g_-,\Psi_{-}]}\,,
\end{align}
where the $\oint$ symbol indicates that: 1) a trace over those degrees of freedom is performed in the far future, and 2) the initial state of those degrees of freedom is accounted for by the boundary conditions in the path-integral, both of which correlate the $\pm$ variables. 

In \cref{eq:SKWLIeff} we've ignored direct dissipative interactions between the particle and environment, e.g. aerodynamic surface drag forces, as we're focusing on the long-ranged gravitationally induced coupling to the environment. These drag forces  are examples of the previously discussed ``finite-size'' effects since the strength of these interactions is set by positive powers of $R$, the size of the object. Mixed interactions involving both dynamical gravity and finite-size interactions with the environment are subleading, and will be treated in a separate paper~\cite{GMRII}.

\subsubsection{The environment}

We can first integrate out the environment to obtain an effective gravitational action. Since we are interested in non-trivial environments we will utilize the background field method. We perturb the metric about flat spacetime, $g_{\pm} = \eta + h_{\pm}$, write the environment variables as $\Psi_{\pm} = \bar{\Psi}+\psi_{\pm}$, and proceed treating $(h,\psi)$ perturbatively since they scale as powers of $\epsilon_{1,2}$.

Since we're omitting the subleading finite-size effects in this work, we can simplify $S_{kin}$ and treat the mass as only depending on the background value of the  fields,
\begin{equation}
    M(\dot{x},\Psi,\mu) = M(\bar{\Psi},\mu)+\mathcal{O}(\psi).
\end{equation}
The mass will also depend on the  renormalization scale, $\mu$, but to improve readability we will omit these arguments and write this simply as $M$ until the final expression.

As in the previous section we will use Keldysh $(r,a)$ variables. Since we're working perturbatively about flat spacetime these are simply averages and differences of the $(+,-)$ variables.
\begin{align}\label{eq:gravEffActionSK}
   e^{iW[h_r,h_a]} &= \oint \mathcal{D}\Psi_+ \mathcal{D}\Psi_-\,\,e^{iS_{env}[\Psi_+,g_+]-iS_{env}[\Psi_-,g_-]} \nonumber \\
   &= \left\langle e^{\frac{-i}{2}\int d^{4}x\left( h_{\mu\nu\,+}T_{+E}^{\mu\nu}-h_{\mu\nu\,-}T^{\mu\nu}_{- E}\right)+O(h^2)} \right\rangle \nonumber \\
   &= \left\langle e^{\frac{-i}{2}\int d^{4}x\left( h_{\mu\nu\,a}T_{rE}^{\mu\nu}+h_{\mu\nu\,r}T^{\mu\nu}_{aE}\right)+O(h^2)} \right\rangle\,
\end{align}
Due to the universality of gravitational interactions we do not need to know, at leading order, any details about the environment other than the correlation function of its energy-momentum tensor. Next we calculate the effective action for the metric  by integrating out fluctuations of the environment around its background,
\begin{align}\label{eq:gravEffActionCumulantExp}
iW[h_r,h_a] &= \frac{-i}{2}\int d^{4}x\, h_{\mu\nu\,a}(x)  \langle T^{\mu\nu}_{rE}(x)\rangle  \nonumber \\
&+\frac{1}{2}\left(\frac{-i}{2}\right)^{2}\int d^{4}x d^{4}y \left[h_{\mu\nu\,a}(x)h_{\sigma\rho\,a}(y)\left\langle \delta T^{\mu\nu}_{rE}(x)\delta T^{\sigma\rho}_{rE}(y)\right\rangle+2h_{\mu\nu\,a}(x)h_{\sigma\rho\,r}(y)\left\langle \delta T^{\mu\nu}_{r E}(x)\delta T^{\sigma\rho}_{a E}(y)\right\rangle\right] \nonumber \\
&+ \mathcal{O}(h^{3})\,.
\end{align}
We use the notation $\delta T^{\mu\nu}$ to denote fluctuations of the environment around its background value $\langle T_{rE}^{\mu \nu} \rangle$. Since we linearized the action in \eqref{eq:gravEffActionSK} one might worry that we could not reliably expand to quadratic order in \eqref{eq:gravEffActionCumulantExp}. Indeed we are omitting the so-called seagull vertex describing the simultaneous coupling of two gravitons to the environment. This coupling cannot describe dissipative losses to the environment, so we will omit it in this work. It will become relevant for conservative forces however, which we will present elsewhere.

At quadratic order there are two distinct types of correlators which appear. A thorough discussion of the $(r,a)$ correlators is given in many references (see e.g. \cite{Liu:2018kfw}), here we will mention only the essential details for two-point correlators. For a given correlator one can unpack the $(r,a)$ variables into $(+,-)$ variables and thereby determine the correlator in terms of standard operator calculus. For example, for an operator $\mathcal{O}(t)$ one has
\begin{align}
    \langle \mathcal{O}_r(t)\rangle &= \langle\mathcal{O}(t)\rangle, 
    \qquad\qquad\langle\mathcal{O}_{r}(t_1)\mathcal{O}_{a}(t_2)\rangle =\theta(t_1-t_2)\langle[\mathcal{O}(t_1),\mathcal{O}(t_2)]\rangle\equiv (-i)G_{ret}(t_1,t_2)\nonumber \\
    \langle \mathcal{O}_a(t)\rangle &= 0 
    \qquad\,\,\;\qquad\qquad\langle\mathcal{O}_{a}(t_1)\mathcal{O}_{r}(t_2)\rangle = -\theta(t_2-t_1)\langle[\mathcal{O}(t_1),\mathcal{O}(t_2)]\rangle\nonumber\equiv  (-i)G_{adv}(t_1,t_2) 
    \nonumber \\
    \langle\mathcal{O}_{a}(t_1)\mathcal{O}_{a}(t_2)\rangle&=0
    \qquad\,\,\;\qquad\qquad\langle\mathcal{O}_{r}(t_1)\mathcal{O}_{r}(t_2)\rangle = \frac{1}{2}\langle \{\mathcal{O}(t_1),\mathcal{O}(t_2)\}\rangle\equiv G_{sym}(t_1,t_2)\,. 
\end{align}
The $ra$ correlator is simply the causal linear response function, and the $rr$ correlator is the noise-kernel describing stochastic/quantum fluctuations.\footnote{For a finite temperature environment these fluctuations are determined by the dissipative part of the linear response function via the fluctuation-dissipation theorem.} While the CTP effective action is sufficiently general to account for environmental fluctuations in this work we will focus only on the thermodynamic limit and omit the stochastic physics as sub-leading to the deterministic physics. We will henceforth drop all $rr$ correlators. It would be interesting future work, however, to understand the effects of stochasticity.

The effective action is generically nonlocal for the homogeneous finite-density media considered here. In the thermodynamic limit, conservation of energy and momentum, together with $\rho_{E}+p_{E}\neq0$, implies gapless long-wavelength collective response, such as hydrodynamic, ballistic, or collisionless modes. Both terms, in the second line in $W$, scale as the environmental mass density $\rho_{E}$, and so each insertion of a vertex from $W$ into a Feynman diagram will be penalized by a small factor of $\epsilon_2$.

\subsubsection{The metric perturbations}

The next step is to integrate out the metric perturbations. To do so, we first want to write the Einstein-Hilbert and particle actions in the Keldysh variables. Since we're only interested in the deterministic equations of motion, neglecting stochastic fluctuations, we only  keep  terms linear  in the $a$-variables. Additionally, since we're computing just the leading order result we can truncate the expansions in metric perturbations to leading order. 

We define the CTP actions,
\begin{align}
   S_{EH}[h_r,h_a] &= S_{EH}[g_+]-S_{EH}[g_-]\,, \nonumber \\
   S_{kin}[x_r,h_r,x_a,h_a] &= S_{kin}[x_+,g_+]-S_{kin}[x_-,g_-]\,.
\end{align}
To leading order in metric perturbations, the Einstein-Hilbert action is, in Lorenz/de Donder gauge,
 \begin{align}\label{eq:EHlinearized}
   S_{EH}[h_r,h_a] = \frac{-1}{32\pi G} \int d^{4}x \,h_{\mu\nu\,a}\,\partial^{2}\bar{h}_{r}^{\mu\nu}\,,
\end{align}
where the overbar denotes trace reversal.

Using the gauge fixing condition 
\begin{equation}
\label{gauge}
    1=g_{\mu\nu}\frac{dx^{\mu}}{d\lambda}\frac{dx^{\nu}}{d\lambda},
\end{equation}
on solutions to the equation of motion, we can write the usual (single time-path) particle action as
\begin{align}\label{eq:ppaction}
    S_{kin}[x,h] = -\frac{M}{2}\int d\lambda \left[(\eta_{\mu\nu}+h_{\mu\nu})\dot{x}^{\mu}\dot{x}^{\nu}\right] = -\frac{1}{2}\int d^{4}y\,T^{\mu\nu}(y)[\eta_{\mu\nu}+h_{\mu\nu}(y)]\,.
\end{align}
The CTP particle action is then
\begin{align}\label{eq:particle-source-response}
    S_{kin}[x_r,h_r,x_a,h_a] = -\frac{1}{2}\int d^{4}y\, T^{\mu\nu}_{r} h_{\mu\nu\,a}-\frac{1}{2}\int d^{4}y\, T^{\mu\nu}_{a} (\eta_{\mu\nu}+h_{\mu\nu\,r})\,,
\end{align}
and to linear order in $a$ variables we simply have
\begin{align}
    T^{\mu\nu}_{r}(y) &= M\int d\lambda\,\dot{x}^{\mu}_r\dot{x}^{\nu}_r\,\delta^{4}(y-x_r(\lambda))\,,\nonumber \\
    T^{\mu\nu}_{a}(y) &= M\int d\lambda \big[\dot{x}_a^{\mu}\dot{x}_r^{\nu}+\dot{x}^{\mu}_r\dot{x}_a^{\nu}-\dot{x}_r^{\mu}\dot{x}_r^{\nu}x^{\rho}_a\partial_{\rho}\big]\delta^{4}(y-x_r(\lambda)).
\end{align}

To evaluate $I_{\mathrm{eff}}[x_r,x_a]$, we  compute all connected tree-level \footnote{Including a line for the world-line propagator gives the false impression that there is a loop involved.} Feynman diagrams.\footnote{Had we kept the stochastic $rr$ correlators then certain classical loop diagrams would remain, as in fluctuating hydrodynamics or the EFT of LSS, but this is of no concern for the present work.} Since the correlators $\langle h_{r}h_{a}\rangle$ are causal, with $h_a$ necessarily to the causal past of $h_r$, we can assign a causal arrow to the graviton propagators. This implies that the $\int T h_{a}$ vertices are sources while the $\int T h_{r}$ vertices are responses.

\subsection{The leading-order diagram}\label{sec:LOdiagram}

Using the ingredients above, we will now integrate out the graviton to obtain the leading order contributions to the point-particle effective action.

\subsubsection{Position space}

The leading diagrams contributing to the particle's effective CTP action are the following. We use the notation $I_{\mathrm{eff}}^{(m,n)}$ to denote order $\epsilon_{1}^{m}\epsilon_{2}^{n}$ contributions. At order $\epsilon_2^{0}$, we have just the gravitational radiation-reaction contribution
\begin{equation}
    I_{\mathrm{eff}}^{(2,0)}[x_r,x_a] = 8\pi G \int d^{4}yd^{4}y' \,T_{a}^{\mu\nu}(y)D_{\mu\nu; \alpha\beta}(y,y')T^{\alpha\beta}_r(y')\,.
\end{equation}
where $D_{\mu\nu; \alpha\beta}(y,y')$ is the retarded Green's function for the metric perturbation in vacuum.\footnote{Since we're expanding $g_{\mu\nu}=\eta_{\mu\nu}+h_{\mu\nu}$, the graviton is not canonically normalized (see \cref{eq:EHlinearized}). As such, the graviton propagator we've used is $(16\pi G)D_{\mu\nu;\alpha\beta}$, where $D_{\mu\nu;\alpha\beta}$ is the canonically normalized Green's function in de Donder gauge.}  The leading quadrupolar energy loss and its secular effect on binary
motion were established in
Refs.~\cite{PetersMathews1963,Peters1964}, while the explicit
\(2.5\)PN radiation-reaction force was derived in
Refs.~\cite{ChandrasekharEsposito1970,BurkeThorne1970,Burke1971}.
The corresponding higher-order post-Newtonian, nonlinear, and
self-force effects are also well understood
\cite{IyerWill1993,IyerWill1995,Blanchet1997,PatiWill2002,
MinoSasakiTanaka1997,QuinnWald1997,GalleyTiglio2009,
GalleyLeibovich2012,GoldbergerRoss2010,GalleyEtAl2016,
LeibovichPardoYang2023,Blanchet2024}, we therefore proceed to
contributions involving the environment.

 At first order in $\epsilon_2$, we have two contributions. The first is the direct gravitational pull that the environment exerts on the particle
\begin{equation}
    I_{\mathrm{eff}}^{(1,1)}[x_r,x_a] = 8\pi G\int d^{4}yd^{4}y' \,T_{a}^{\mu\nu}(y)D_{\mu\nu;\alpha\beta}(y,y') \langle T_E^{\alpha\beta}(y')\rangle\,,
\end{equation}
which  exerts no dissipative force on the particle and we can ignore this contribution. 

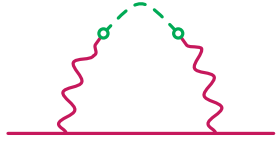
\begin{figure}[H]
\centering

\begin{tikzpicture}[line cap=round, line join=round, thick, scale=0.55]

  \definecolor{magentaLine}{RGB}{198, 24, 92} 
  \definecolor{greenArc}{RGB}{0, 170, 85}     

  \tikzset{
    wavy/.style={
      draw=magentaLine,
      very thick,
      decorate,
      decoration={snake, amplitude=1.0mm, segment length=4.0mm}
    },
    base/.style={draw=magentaLine, very thick},
    arc/.style={draw=greenArc, very thick, dashed, dash pattern=on 4pt off 4pt}
  }

  \coordinate (L)  at (-1.8,0);
  \coordinate (R)  at ( 1.8,0);
  \coordinate (LT) at (-0.9,2.4);
  \coordinate (RT) at ( 0.9,2.4);

  \draw[base] (-3.2,0) -- (3.2,0);

  \draw[wavy] (L) .. controls (-1.8,0.9) and (-1.35,1.8) .. (LT);
  \draw[wavy] (R) .. controls ( 1.8,0.9) and ( 1.35,1.8) .. (RT);

  \draw[arc] (LT) .. controls (0,3.35) .. (RT);

  \draw[greenArc, very thick, fill=white] (LT) circle (0.11);
  \draw[greenArc, very thick, fill=white] (RT) circle (0.11);

\end{tikzpicture}

\caption{Leading order contribution to the effective action. The wavy lines are graviton propagators, and the dashed line is the stress-energy two-point function. In the presence of an environment the graviton mixes with phonons and other light environmental modes.}
\label{fig:my-diagram}

\end{figure}

At $\mathcal{O}(G^2)$ we have the diagram of interest, \cref{fig:my-diagram}, involving the linear response of the environment
\begin{equation}
    I_{\mathrm{eff}}^{(2,1)}[x_r,x_a] = (8\pi G)^2\int d^{4}yd^{4}y' \,T_{a}^{\mu\nu}(y)G^{R}_{\mu\nu\alpha\beta}(y,y') T^{\alpha\beta}_{r}(y')\,,
\end{equation}
where the effective graviton Green's function is
\begin{equation}\label{eq:retarded_green_function}
    G^{R}_{\mu\nu\alpha\beta}(y,y') = \int d^{4}z d^{4}z'\,D_{\mu\nu;\rho\sigma}(y,z)\langle\delta T_E^{\rho\sigma}(z)\delta T_E^{\gamma\delta}(z')\rangle^{ret} D_{\gamma\delta;\alpha\beta}(z',y')\,.
\end{equation}
At $\mathcal{O}(G^{2})$  the seagull vertex also contributes but does not lead to any   dissipative force on the object, and so we omit it here. 

We emphasize that because the coupling to the graviton is universal we did not need to know any details of the bulk EFT describing the environment other than its stress-energy linear response function. Additionally, because the whole construction is based on a CTP contour this response function can be intrinsically dissipative, as is the case for a viscous fluid, for example.

To develop intuition for these results, it is helpful to recognize that the ``response'' vertex in \eqref{eq:particle-source-response} (i.e. the vertex involving $T^{\mu\nu}_{a}$) has an identical form to the standard point-particle action \eqref{eq:ppaction}. Given this, and the fact that our diagrams necessarily have only one response vertex, we can understand the effective action as simply
\begin{equation}
    I_{\mathrm{eff}}[x_r,x_a] = \left(x^{\mu}_{a}*\frac{\delta}{\delta x^{\mu}_r}\right) S_{kin}[x_r,\eta+ \langle h_r\rangle]\,,
\end{equation}
where $\langle h_{\mu\nu\,r}\rangle$ is the metric perturbation sourced by both environment and the particle itself and
$*$ means integrated over affine parameter. The equations of motion follow by varying $I_{\mathrm{eff}}[x_r,x_a]$ with respect to $x_{a}^{\mu}$, which we can immediately see is equivalent to varying $S_{p}[x_r,\eta+ \langle h_r\rangle]$ with respect to $x_r$. In the absence of external forces, the equation of motion for the point particle is then simply the geodesic equation on the metric $g_{\mu\nu}=\eta_{\mu\nu}+ \langle h_{\mu\nu\,r}\rangle$. 

From the leading diagrams computed here, we can immediately read off the contributions to the effective metric perturbation
\begin{align}
    h^{(1,0)}_{\mu\nu}(y) &= -16\pi GM\int d\lambda' D_{\mu\nu;\alpha\beta}(y,x(\lambda'))\,v^{\alpha}(\lambda')v^{\beta}(\lambda')\,, \label{eq:h20}\\
    h^{(0,1)}_{\mu\nu}(y) &= -16\pi G\int d^{4}y' D_{\mu\nu;\alpha\beta}(y,y') T_E^{\alpha\beta}(y')\,,  \label{eq:h11}\\
    h^{(1,1)}_{\mu\nu}(y) &= -\frac{1}{2}(16\pi G)^2M\int d\lambda' G^{R}_{\mu\nu;\alpha\beta}(y,x(\lambda'))\,v^{\alpha}(\lambda')v^{\beta}(\lambda')\,.\label{eq:h21}
\end{align}
As there should be no more risk of confusion we have dropped the angled brackets and the subscripts, $x=x_{r}$ and $h=h_r$. We have also introduced the particle's four-velocity, $v^{\mu}=\dot{x}^{\mu}$.

We generate the force law by varying the effective action to obtain the equation of motion,
\begin{equation}\label{eq:correctedEOM}
M \frac{d v^\mu}{d\lambda}
=
F^\mu_{\rm ext}
-
M\,\delta\Gamma^\mu_{\alpha\beta} v^\alpha v^\beta
-
h^{\mu\nu}F^{\rm ext}_\nu
+\cdots .
\end{equation}
This is the force as parameterized by the full metric proper time (see  eq. \ref{gauge}). However, this is not the most convenient gauge since the affine parameter will then implicitly depend upon $h$.
So for bookkeeping purposes it proves more useful
to parameterize by the flat-space (non-affine) proper time
such that $\dot x^\mu \dot x^\nu \eta_{\mu \nu}=1$.
For such a non-affine parameter the (force-free) geodesic equation is
\begin{equation}
\ddot{x}^{\mu}
+
\Gamma^{\mu}_{\alpha\beta}\dot{x}^{\alpha}\dot{x}^{\beta}
=
\frac{\dot e}{e}\dot{x}^{\mu}\,,
\end{equation}    
where $e= \sqrt{\dot x^2}$. At linearized order $\frac{\dot e}{e}=\frac{1}{2}v\cdot \partial h_{\mu \nu}v^\mu v^\nu=
v_\mu \delta\Gamma^\mu_{\alpha\beta}v^\alpha v^\beta
+O(h^2).$
The force free geodesic equation can then be written as 
\begin{equation}
\frac{d v^\mu}{d\tau}
=
-
P^{\mu\nu}
\delta\Gamma_{\nu\alpha\beta}
v^\alpha v^\beta ,
\qquad
P^{\mu\nu}
\equiv
\eta^{\mu\nu}-v^\mu v^\nu .
\end{equation}

Thus the effect of the  transformation simply introduces the orthogonal projector into our force law~\cite{Poisson:2011nh}.
The  external force is defined
as the force on a point mass in flat space, and so is orthogonal to the velocity.
This is not an assumption but a definition, as the action of the force should preserve the mass shell condition.
Repristinating the mass  and projecting the geodesic equation onto the velocity gives
\begin{equation}
M\left(
v_\mu \delta\Gamma^\mu_{\alpha\beta}v^\alpha v^\beta
-
\frac{\dot e}{e}
\right)
=
v_\mu F^\mu_{\rm ext}
-
v_\mu h^{\mu\nu}F^{\rm ext}_\nu,
\end{equation}
where the einbein is $e=\sqrt{\dot x^2}$.
With the external force we have
\begin{equation}
\frac{\dot e}{e}
=
v_\mu\delta\Gamma^\mu_{\alpha\beta}v^\alpha v^\beta
+
\frac{1}{M}
v_\mu h^{\mu\nu}F^{\rm ext}_\nu .
\end{equation}
Plugging this back into the geodesic equation we have
\begin{equation}
M\frac{dv^\mu}{d\tau}
=
F^\mu_{\rm ext}
-
M P^{\mu\nu}
\delta\Gamma_{\nu\alpha\beta}v^\alpha v^\beta
-
P^\mu{}_\rho h^{\rho\nu}F^{\rm ext}_\nu .
\end{equation}

 Thus gravitational self-force in an environment has a universal leading order description when computed perturbatively about flat spacetime. We summarize the result in the following set of equations
\begin{align}\label{eq:forcelawsummary}
    M\frac{dv^{\mu}}{d\tau} &= F_{ext}^{\mu}(x)+F_{DF}^{\mu}(x)\,,\nonumber \\
    F_{DF}^{\mu}(x) &=  -M P^{\mu\nu}v^{\alpha}v^{\beta}\delta\Gamma_{\nu\alpha\beta}-P^\mu_\rho h^{\rho\nu}F_{ext\,\nu}\,,\nonumber \\
    \delta\Gamma_{\nu\alpha\beta} &= \frac{1}{2}\left(\partial_{\alpha}h_{\beta\nu}+\partial_{\beta}h_{\alpha\nu}-\partial_{\nu}h_{\alpha\beta}\right)\,,\nonumber \\
    h_{\mu\nu}(y) &= -\frac{1}{2}(16\pi G)^2M\int d\tau' G^{R}_{\mu\nu;\alpha\beta}(y,x(\tau'))\,v^{\alpha}(\tau')v^{\beta}(\tau')\,,\nonumber \\
    G^{R}_{\mu\nu\alpha\beta}(y,y') &= \int d^{4}z d^{4}z'\,D_{\mu\nu;\rho\sigma}(y,z)\langle\delta T^{\rho\sigma}(z)\delta T^{\gamma\delta}(z')\rangle^{ret} D_{\gamma\delta;\alpha\beta}(z',y')\,,
\end{align}
where $P^{\mu \nu}=(\eta^{\mu \nu}-v^\mu v^\nu)$ is the transverse projector  and $F_{ext}^{\mu}$ is the zeroth-order externally applied force and $F_{DF}^{\mu}$ the leading environmentally dressed self-force, $D_{\gamma\delta;\alpha\beta}$  the retarded Green's function for metric perturbations and $\langle\delta T^{\rho\sigma}(z)\delta T^{\gamma\delta}(z')\rangle^{ret}$ being the retarded linear-response function for stress-energy perturbations of the environment. The gravitational radiation-reaction force and the stationary gravitational attraction of the environment can  readily be included by adding \cref{eq:h20,eq:h11} to  $h_{\alpha\beta}$.

In the above expressions $M$ is the effective renormalized
mass that depends upon the choice of environment as well as a the choice of boundary conditions. The force $F_{DF}$ has both a conservative  and  dissipative piece. The dissipative piece will be independent of the choice of gauge for the graviton. The conservative force will generically depend on the gauge, as forces are obviously frame-dependent notions. Furthermore, we've omitted two diagrams which may contribute to the relativistic conservative force so the conservative part of \cref{eq:forcelawsummary} can only be trusted in the non-relativistic limit.

The conservative part of the force will vanish for highly symmetric backgrounds/trajectories. More generally, it does not lead to secular decay like the dissipative piece, but does cause wave dephasing. We will return to the
conservative force in a forthcoming publication.

\subsubsection{Fourier space}

Let us focus on the case of a homogeneous and isotropic environment. In this case we can dramatically simplify the force law in \eqref{eq:forcelawsummary} via Fourier transform. 

To repackage the index structure in the Christoffel symbol it is convenient to define
\begin{equation}
    \mathcal{O}_{\nu}^{\,\,\alpha\beta}(v,\partial) = \frac{1}{2}(\eta^{\alpha}_{\nu}v^{\beta}+\eta^{\beta}_{\nu}v^{\alpha})(v\cdot\partial)-\frac{1}{2}v^{\alpha}v^{\beta}\partial_{\nu}\,.
\end{equation}
so that the force on the particle at $x^{\mu}(\tau)$ can be written
\begin{equation}
    F^{\mu}(x(\tau))= -MP^{\mu\nu}\mathcal{O}_{\nu}^{\,\,\alpha\beta}\big(v(\tau),\partial\big)h_{\alpha\beta}(x(\tau))\,.
\end{equation}
Here, and in the rest of paper, we focus on situations with no external forces.

Inserting the metric perturbation in \eqref{eq:forcelawsummary}, we obtain the one-loop momentum space integral which computes the force
\begin{equation}\label{eq:forcetailintegral}
    F^{\mu}= -i\frac{\kappa^4 M^2}{8}(\eta^{\mu\nu}-v^{\mu}v^{\nu})P_{\alpha\beta ab}P_{cd\rho\sigma} \int^{\infty}_{\tau_{0}} d\tau' v^{\rho}(\tau')v^{\sigma}(\tau') \int_{k} e^{-ik\cdot x(\tau)+ik\cdot x(\tau')} \frac{\mathcal{O}_{\nu}^{\,\,\alpha\beta}(v,k)}{k^4}\langle\delta T^{ab}\delta T^{cd}\rangle^{ret}(k)\,,
\end{equation}
where $\kappa^2=32\pi G$, $\tau_{0}$ is proper time at which the interactions started, and unless otherwise specified $v=v(\tau)$.  In this expression we've chosen the  de Donder gauge, hence the appearance of the trace reversal operator
\begin{equation}
    P_{\alpha\beta ab}=\frac{1}{2}\left(\eta_{\alpha a}\eta_{\beta b}+\eta_{\alpha b}\eta_{\beta a}-\eta_{\alpha\beta}\eta_{ab}\right)\,.
\end{equation}

As discussed in the introduction, this force integral may, or may not,  depend on the past history of the trajectory depending upon the nature of the response function \footnote{If the response can be reliably expanded in time differences then the system will be Markovian.}.  The evaluation of this ``tail'' integral is then very challenging in general. In  \cref{sec:SFingeneralenv} we will revisit this expression for a rectilinear trajectory, along which the integral can be evaluated.

To isolate the dissipative part of the force we consider the  time reversal asymmetric contribution. In momentum space this piece will pick out the piece  of the stress-energy correlator which is odd under $k\to -k$. Since the stress-energy tensor is hermitian, we have
\begin{equation}
    \langle\delta T^{\mu\nu}\delta T^{\alpha\beta}\rangle^{ret}(-k) =     \big[\langle\delta T^{\mu\nu}\delta T^{\alpha\beta}\rangle^{ret}(k)\big]^{*}\,, 
\end{equation}
and the part of the correlator which is odd in $k$ is just the imaginary part, otherwise known as the spectral density
\begin{equation}\label{eq:stress_tensor_spectral_density}
   \sigma_{T}^{\mu\nu\alpha\beta}(k) \equiv 2\Im  \langle\delta T^{\mu\nu}\delta T^{\alpha\beta}\rangle^{ret}(k)  = -i\left[\langle\delta T^{\mu\nu}\delta T^{\alpha\beta}\rangle^{ret}(k)-\langle\delta T^{\mu\nu}\delta T^{\alpha\beta}\rangle^{ret}(-k)\right]\,.
\end{equation}
and the dissipative force is then
\begin{equation}\label{eq:dissipforce}
    F_{diss}^{\mu}= \frac{\kappa^4 M^2}{16}
    (\eta^{\mu\nu}-v^{\mu}v^{\nu})
    P_{\alpha\beta ab}P_{cd\rho\sigma}
    \int^{\infty}_{\tau_{0}} d\tau' v^{\rho}(\tau')v^{\sigma}(\tau')
    \int_{k}e^{-ik\cdot(x(\tau)-x(\tau'))}
    \frac{\mathcal{O}_{\nu}^{\,\,\alpha\beta}(v,k)}{k^4}
    \sigma_{T}^{abcd}(k)
\end{equation}

If we were interested in the contribution to the energy loss from gravitational radiation then we would need to worry about the graviton radiation poles which describe environmental dressing effects on the propagation of gravitational (e.g. Burke-Thorne) radiation and is subleading in a PN expansion.

\subsection{The general form of stress-energy correlators}

We will assume a  homogeneous and isotropic environment,  which implies  that the only distinguished vector is the four-velocity which picks out the environment's rest frame, $u^{\mu}$. Lorentz symmetry  constrains the equilibrium stress-tensor to have the isotropic fluid form
\begin{equation}\label{eq:fluidstresstensor}
    \langle T^{\mu\nu}\rangle = \rho_{E} u^{\mu}u^{\nu}+p_{E}(u^{\mu}u^{\nu}-\eta^{\mu\nu})\,,
\end{equation}
and from thermodynamic considerations one understands the constants $\rho_{E},p_{E}$ to be the energy density and pressure.

\subsubsection{Tensor Decomposition}\label{sec:tensordecomp}

Given our assumption of homogeneity, $k^\mu$ and $u^\mu$ will break the local Lorentz group
down to the little group of residual transformations $SO(D-2)$. That the group is Euclidean follows from the fact that
$u^\mu$ is timelike and then $k$ breaks the residual $SO(D-1)$ down to $SO(D-2)$. The tensor structure must
therefore be invariant under the $SO(D-2)$. We can define the direction orthogonal to $k$
\begin{equation}
    l^{\mu}=(u\cdot k) k^{\mu} - k^2 u^{\mu}\,,
\end{equation}
which is the projection of $u$ onto the direction orthogonal to $k$.

Now we would like to decompose the product of two symmetric tensors into irreps of the little group. We start by first 
decomposing one symmetric tensor into the full space and then use those decompositions to find invariants
in the product. We can write the full space as $Span(k,l,n)$ where $n$ is the space orthogonal to $k$ and $l$.
We wish to form all possible symmetric rank two tensors. There are seven such tensors: Four scalars $k^\mu k^\nu, l^\mu l^\nu$ and $k^{(\mu} l^{\nu)}$ and $\Pi^{\mu \nu}$. Where $\Pi^{\mu \nu}$ is the projector
onto the orthogonal subspace 
\begin{equation}
\Pi^{\mu\nu}
\equiv
\frac{l^\mu l^\nu}{l^2}
+
\frac{k^\mu k^\nu}{k^2}
-
\eta^{\mu\nu}.
\end{equation}
There are two vectors $k^{(\mu}X_\perp^{\nu)}$, and
$l^{(\mu}X_\perp^{\nu)}$  and one symmetric trace free tensor $(X^\mu_\perp X^\nu_{\perp})_{STF}$.

Now we can decompose $T^{\mu \nu}$ as follows
\begin{equation}
\begin{aligned}
T^{\mu\nu}
={}&
A\,k^\mu k^\nu
+B\,l^\mu l^\nu
+C\left(k^\mu l^\nu+l^\mu k^\nu\right)
+D\,\Pi^{\mu\nu}
\\
&+
k^\mu A_\perp^\nu+k^\nu A_\perp^\mu
+l^\mu B_\perp^\nu+l^\nu B_\perp^\mu
+C_\perp^{\mu\nu}.
\end{aligned}
\end{equation}

Then imposing transversality this reduces to
\begin{equation}
T^{\mu\nu}
=
B\,l^\mu l^\nu
+
D\,\Pi^{\mu\nu}
+
l^\mu B_\perp^\nu+l^\nu B_\perp^\mu
+
C_\perp^{\mu\nu}.
\end{equation}
We have kept non-invariants since we will contract them to form invariants when we
consider the correlator.

First we have the two-index tensors which are little group scalars, defining
\begin{equation}
    S^{\mu\nu}_{0}=\frac{l^{\mu}l^{\nu}}{l^2},\qquad S^{\mu\nu}_{1}=\Pi^{\mu\nu}\,.
\end{equation}
Thus the  scalar contribution  to the correlator can be written in terms of the three independent tensor structures
\begin{align}\label{eq:scalarformfactordecomposition}
    \mathcal{S}_{00}^{\mu\nu\alpha\beta}&=S_{0}^{\mu\nu}S_{0}^{\alpha\beta} \nonumber \\
    \mathcal{S}_{11}^{\mu\nu\alpha\beta}&=S_{1}^{\mu\nu}S_{1}^{\alpha\beta} \nonumber \\
    \mathcal{S}_{01}^{\mu\nu\alpha\beta}&=S_{0}^{\mu\nu}S_{1}^{\alpha\beta} +S_{1}^{\mu\nu}S_{0}^{\alpha\beta}\,.
\end{align}
For the vector piece the correlator must take the form
\begin{equation}
\left\langle
B_\perp^\mu(k)\,
B_\perp^\nu(-k)
\right\rangle
=
G_V\!\left(k^2,u\cdot k\right)\,
\Pi^{\mu\nu}.
\end{equation}
since $\Pi^{\mu \nu}$ is the only singlet  that lives in the orthogonal subspace.
Then the vector contribution to the two-point function can be parameterized by
\begin{equation}
\begin{aligned}
\mathcal{V}^{\mu\nu\alpha\beta}
=
\frac{1}{l^2}
\Big(
&
l^\mu l^\alpha \Pi^{\nu\beta}
+
l^\mu l^\beta \Pi^{\nu\alpha}
+
l^\nu l^\alpha \Pi^{\mu\beta}
+
l^\nu l^\beta \Pi^{\mu\alpha}
\Big).
\end{aligned}
\end{equation}

Finally we can build the symmetric traceless representation of the little group via
\begin{equation}
    \mathcal{T}^{\mu\nu\alpha\beta}
    =
    \frac{1}{2}\Pi^{\mu\alpha}\Pi^{\nu\beta}
    +\frac{1}{2}\Pi^{\mu\beta}\Pi^{\nu\alpha}
    -\frac{1}{D-2}\Pi^{\mu\nu}\Pi^{\alpha\beta}.
\end{equation}

The general tensor decomposition of the stress-energy tensor spectral density is given by linear combinations of the above irreducible representations
\begin{equation}\label{eq:formfactordecomposition}
    \sigma_{T}^{\mu\nu\alpha\beta}(k) = G_{00}(k) \mathcal{S}_{00}^{\mu\nu\alpha\beta} + G_{11}(k)\mathcal{S}_{11}^{\mu\nu\alpha\beta}+G_{01}(k)\mathcal{S}_{01}^{\mu\nu\alpha\beta}+G_{V}(k)\mathcal{V}^{\mu\nu\alpha\beta}+G_{\tau}(k)\mathcal{T}^{\mu\nu \alpha\beta}\,.
\end{equation}

 The above form factors can be compared to conventional hydrodynamic modes by fixing a rest frame $u^{\mu}=(1,0,...,0)$ and a perturbation wave vector along the z-axis, $k^{\mu}=(\omega, 0,...,0,|\vec{k}|)$. We find:
\begin{align}\label{eq:transverseFormFactors}
    G_{00}(k) &= \frac{k^4}{|\vec{k}|^4}2\Im\langle\mathcal{E}\mathcal{E}\rangle\nonumber \\
    G_{11}(k) &= 2\Im\langle P_{\perp} P_{\perp}\rangle  \nonumber \\
    G_{01}(k) &= -\frac{k^2}{|\vec{k}|^2}2\Im\langle\mathcal{E} P_{\perp}\rangle \nonumber \\
    G_{V}(k) &= -\frac{k^2}{|\vec{k}|^2}2\Im\langle\pi_{\perp}\pi_{\perp}\rangle\nonumber \\
    G_{\tau}(k) &= 4\Im\langle\tau_{\perp}\tau_{\perp}\rangle\,,
\end{align}
where we've defined perturbations of the energy density, transverse pressure, transverse momentum, and  shear stress,
\begin{align}
    \mathcal{E} &= \delta T^{00}\,, \nonumber \\
    P_{\perp} &= \frac{\delta_{AB}}{D-2}\delta T^{AB}\,, \nonumber \\
    \pi_{\perp} &= \delta T^{0x}\,, \nonumber \\
    \tau_{\perp} &= \delta T^{xy}\,.
\end{align}
and $A,B$ run over the $D-2$ spatial directions orthogonal to $u^{\mu},k^{\mu}$, (i.e. orthogonal to the $0$ and  $z$ axes in this frame).

\subsection{Dynamical friction in general environments}\label{sec:SFingeneralenv}

The environmentally induced self-force depends on the history of the particle's trajectory, and under the influence of this self-force the trajectory of the particle will evolve. However, because the force is perturbatively small the trajectory will evolve slowly. In a leading adiabatic approximation one can then treat $(x^{\mu}(\tau'),v^{\mu}(\tau'))$ in the force integral as being determined solely by the zeroth-order dynamics. For straight-line motion, this implies $v^{\mu}(\tau')$ is constant in time.

Our interest will be in the steady-state force, long after the disturbance is beyond the boundary of the environment and transient effects have settled. This is obtained by taking $\tau_{0}\rightarrow-\infty$. Using our formalism there is no obstruction to computing the force in the short-time regime (where environmental disturbances are far from reaching the boundary) first studied in \cite{Ostriker1999}, however, due to transient effects the force is not purely dissipative and additional diagrams contribute.

For uniform motion in steady state the force in \cref{eq:dissipforce} can be written as
\begin{equation}
    F_{diss}^{\mu}= \frac{\kappa^4 M^2}{16}
    (\eta^{\mu\nu}-v^{\mu}v^{\nu})
    P_{\alpha\beta ab}P_{cd\rho\sigma}v^{\rho}v^{\sigma}
    \int^{\infty}_{-\infty} d\tau' 
    \int_{k}e^{-ik\cdot v(\tau-\tau')}
    \frac{\mathcal{O}_{\nu}^{\,\,\alpha\beta}(v,k)}{k^4}
    \sigma_{T}^{abcd}(k)
\end{equation}
The symmetries of the problem ensure that the spatial part of the force is opposite the velocity vector.
By covariance the force must be  proportional to a combination of $u^\mu$ and $v^\mu$. Then
since $v_{\mu}F^{\mu}$ must vanish, the only non-zero contraction with the force is $\dot{E}=-u_{\mu}F^{\mu}$, which has the interpretation of the energy radiated from the particle to the environment per unit proper time, i.e. the energy flux. The force vector can be trivially reconstructed from the energy flux,
\begin{equation}\label{eq:forceFromFlux}
    F_{diss}^{\mu}=-\left(\frac{\gamma v^{\mu}-u^{\mu}}{\gamma^2-1}\right) \dot{E}\,,
\end{equation}
where $\gamma = u\cdot v$.
 We can immediately evaluate the time integral
\begin{equation}
    \int^{\infty}_{-\infty}d\tau' e^{-ik\cdot v(\tau-\tau')} = 2\pi  \delta(k\cdot v)\,.
\end{equation}
On the support of this delta function the various index contractions simplify 
\begin{equation}
    (u^{\nu}-\gamma v^{\nu})\mathcal{O}_{\nu}^{\,\,\alpha\beta}(v,k) = -\frac{1}{2}(k\cdot u) v^{\alpha}v^{\beta}+\frac{1}{2}\underset{=0}{\cancel{(k\cdot v)}} \left(v^{\alpha}u^{\beta}+u^{\alpha}v^{\beta}-\gamma v^{\alpha}v^{\beta}\right)\,.
\end{equation}
For a general translation-invariant environment the steady-state energy flux can then be written  as
\begin{equation}\label{eq:stdystateflux}
    \dot{E}= \frac{\kappa^4 M^2}{32}v^{\mu}v^{\nu}v^{\alpha}v^{\beta}P_{\mu\nu ab}P_{cd\alpha\beta} \int_{k} 2\pi{\delta}(k\cdot v) \frac{k\cdot u}{k^4} \sigma_{T}^{abcd}(k)\,,
\end{equation}

Using the general tensor decomposition, \cref{eq:formfactordecomposition} we obtain
\begin{align}\label{eq:generalsystemformfactedot}
    \dot{E}=\frac{\kappa^4 M^2}{16}\textrm{Im }\int_{k}2\pi\delta(k\cdot v)\,\frac{\omega}{|\vec k|^8}
    \Big[& \langle \mathcal{E}\mathcal{E}\rangle
    \frac{(|\vec k|^2+2k^2\gamma^2)^2}{4}
    -\langle \mathcal{E} P_{\perp}\rangle\gamma^2|\vec k|^2(|\vec k|^2+2k^2\gamma^2)
    +\langle P_{\perp}P_{\perp}\rangle\gamma^4 |\vec k|^4
     \nonumber \\
     &-\langle \pi_{\perp}\pi_{\perp}\rangle 4\gamma^2 |\vec k|^2(|\vec k|^2+k^2\gamma^2)+\langle \tau_{\perp}\tau_{\perp}\rangle \frac{|\vec k|^4(|\vec k|^2+k^2\gamma^2)^2}{k^4}
    \Big]\,,
\end{align}
where  $\omega = u\cdot k$, and $|\vec k|^2 = \omega^2-k^2 $, i.e., in the rest frame of the environment $k^{\mu}=(\omega, \vec{k})$.

This expression holds for any homogeneous and isotropic system, even when a hydrodynamic approximation to the correlators is not valid. For example, one could readily insert correlation functions computed in a weakly coupled quantum field theory. It remains only to insert the imaginary part of these various response functions, i.e. the spectral density in each of the different spin channels.

Equation (\ref{eq:generalsystemformfactedot})  simplifies for many systems of interest. In particular, for systems whose dissipative response lies entirely in the scalar sector and whose scalar anisotropic stress vanishes. This includes perfect fluids, cold dust, and the homogeneous coherent scalar considered below.\footnote{Exceptions include viscous hydrodynamics, or collisionless matter with general velocity-dispersions.}  Stress-energy conservation then gives
\begin{equation}\label{eq:simplecondition}
    P_{\perp} = \delta T^{zz} = \frac{\omega^{2}}{|\vec k|^2}\delta T^{00}.
\end{equation}
then using \cref{eq:generalsystemformfactedot} we find that the sum of contributions from the three scalar modes simplifies considerably 
\begin{align}\label{eq:scalarmodeDF}
    \dot{E}=16\pi^2 G^2M^2(2\gamma^2-1)^2 \int_{k}2\pi\delta(k\cdot v)\,\frac{\omega}{|\vec k|^4}
    \,\,\textrm{Im}\langle \mathcal{E}\mathcal{E}\rangle\,.
\end{align}
All of the post-Minkowskian relativistic effects are captured by the $(2\gamma^2-1)^2$ prefactor.  By explicitly evaluating this integral for various environments we find agreement with results in the literature \cite{Petrich:1989, Barausse:2007ph, Hui:2016ltb}. We will return to this expression in \cref{sec:fluidexample}.

\section{The Universal UV Divergence and the Associated RG Flow}\label{sec:UVandRG}

Working in the EFT illuminates a point that has perhaps been overlooked in the literature on dynamical friction. Namely that the famous ``Coulomb Log'' that arises in Chandrasekhar's calculation is a consequence of the UV divergence in the  one-loop diagram  shown in Fig.~\ref{fig:my-diagram}. Given that this is a short-time effect this divergence will be independent of the path. We may ask when is the log also independent of the choice of environment?

The equivalence principle assures us that the coupling is universal, so
one might expect the short-distance logarithm to be environment independent. The force for a general trajectory is given in \cref{eq:dissipforce}. Thus, the dissipative logarithm is controlled by the high-momentum behavior of $\sigma^{\mu\nu\alpha\beta}_T(k)$. Since UV divergences are local, we  isolate the UV log by expanding the integrand for short times by writing  $x^{\mu}(\tau') \rightarrow x^{\mu}(\tau) + v^{\mu}(\tau)(\tau'-\tau)$ and $v^{\mu}(\tau')\rightarrow v(\tau)$, and integrate over the time parameter. With this replacement the integrand is identical to the steady-state case already considered and the UV divergent part of the energy flux is given by \cref{eq:stdystateflux}, with the instantaneous friction force reconstructed via \cref{eq:forceFromFlux}, and with $v^{\mu}$ understood to be the $\tau$-dependent instantaneous velocity.

We will now focus on the single form factor $\sigma_{\mathcal{E}}\equiv2\Im\langle \mathcal{E}\mathcal{E}\rangle$. As shown in \cref{eq:scalarmodeDF}, for simple environments the relativistic dynamical friction is determined entirely by this form factor. For more general environments, this form factor still controls the leading non-relativistic dynamical friction. 

Since the graviton will mix with any choice of matter, the leading stress tensor will be linear in the matter fields. As such, the leading order (in $G$) stress-energy spectral density will only have delta function support (single particle pole).\footnote{This holds in the absence of intrinsic dissipative behavior, e.g. viscous effects, in the environment. Viscosity is discussed in \cref{eq:viscoussystems}, and is not expected to change the conclusions of this section.} We've assumed that the matter is translation invariant, but breaks Lorentz boosts, so the delta functions can only be functions $\omega  = u\cdot k$ and $|\vec k|^2 = \omega^2-k^2$.  

The form factor is constrained to satisfy the classical version of the f-sum rule
\begin{equation}
\int_{-\infty}^{\infty}\frac{d\omega}{2\pi}\,
\omega\,\sigma_\mathcal{E}(\omega,\vec{k})
=
\left\langle
\left\{
\left\{
H,T^{00}(\vec k,t)
\right\},
T^{00}(-\vec k ,t)
\right\}
\right\rangle_{\rm med} ,
\end{equation}
the right hand side is time independent.
 $\langle\cdots\rangle_{\rm med}$ denotes evaluation in the classical background state of the medium. For a homogeneous stationary medium this average is trivial, in the sense that it only fixes the constant background state and introduces no additional position dependence.
Then stress-energy conservation 
\begin{equation}
\{H,T^{00}\}
= k_i T^{0i} .
\end{equation}
leads to the constraint 
\begin{equation}
\label{eq:constraint_sum_rule}
\int_{-\infty}^{\infty}
\frac{d\omega}{2\pi}\,
\omega\,\sigma_\mathcal{E}
=
 k_i
\left\langle
\left\{
T^{0i},
T^{00}
\right\}
\right\rangle_{\rm med} .
\end{equation}
Assuming a non-relativistic system, we may evaluate the bracket using the Galilean algebra, leaving the constraint 
\begin{equation}
\int_{-\infty}^{\infty}
\frac{d\omega}{2\pi}\,
\omega\,\sigma_\mathcal{E}
=
|\vec k|^2\,\rho_E \,,
\label{eq:nr_f_sum_rule}
\end{equation}
 where \(\rho_E\) is the homogeneous non-relativistic mass density.

The universal non-relativistic dynamical friction energy-flux formula is
\begin{align}
    \dot{E}=8\pi^2 G^2M^2\int_{k}2\pi\delta(\omega - \vec{k}\cdot\vec{v})\,\frac{\omega}{|\vec k|^4}
    \,\,\sigma_{\mathcal{E}}(\omega,|\vec k|)\,,
\end{align}
where in the fluid rest frame, $k^{\mu}=(\omega,\vec{k})$, $v^{\mu}=\gamma(1,\vec{v})$. Performing the angular integration leaves 
\begin{equation}
    \dot{E}=\frac{4\pi G^2M^2}{|\vec v|}\int_{\Lambda_{IR}}^{\Lambda_{UV}}\frac{dk}{k^3} \int \frac{d\omega}{2\pi} \theta(|\vec v|k-|\omega|) \omega \sigma_{\mathcal{E}}(\omega,k)\,.
\end{equation}
The step function will be automatically satisfied provided that $\sigma_\mathcal{E}$  is supported for $|\omega| < |\vec v|k$, as is the case, for example, for supersonic motion in a fluid. The $\omega$ integral is then determined entirely by the sum rule~\eqref{eq:nr_f_sum_rule}, and we obtain the universal result for the divergent part of the Newtonian dynamical friction force
\begin{equation}
    \dot{E}=\frac{4\pi \rho_{E}G^2M^2}{|\vec v|}\int_{\Lambda_{IR}}^{\Lambda_{UV}}\frac{dk}{k}\,.
\end{equation}

We have thus been able to answer the question posed at the beginning of this section in the affirmative, assuming a Newtonian limit.
This answer also assumes that any new scale that may appear in the description of the matter is at much shorter distances than the size of the particle, $R$, which serves as the short-distance cutoff of the EFT. For example,
for a viscous fluid, the viscosity introduces a correction to the dispersion relation that scales as $i k l_{\mathrm{mfp}}$ relative to the real terms, so its effects are suppressed by $l_{\mathrm{mfp}}/R$.

\section{Applications}\label{sec:fluidexample}

\subsection{The inviscid fluid}

As an example application of our general framework, we will recover the  well-known expression for dynamical friction on a body moving relativistically through an ideal fluid. The ideal hydrodynamics linear response functions can be found in many places in the literature, e.g. \cite{Kovtun:2012rj}\footnote{Note, however, the difference in overall sign convention.}.  It is straightforward to verify that they satisfy our ``simple'' condition ($P_{\perp}=\delta T^{zz}$ inside the correlators) so that the dynamical friction is described entirely by \cref{eq:scalarmodeDF}. 

The non-analytic part of the retarded energy density correlator is
\begin{equation}
\langle\mathcal{E}\mathcal{E}\rangle = \frac{(\rho_{E}+p_{E})|\vec k|^2}{c_{s}^2|\vec k|^2-(\omega+i0)^2}\,
\end{equation}
and the imaginary part follows immediately,
\begin{equation}
  \Im \langle\mathcal{E}\mathcal{E}\rangle = \pi(\rho_{E}+p_{E}) |\vec k|^2 \text{sgn}(\omega) \delta(\omega^2-c_s^2|\vec k|^2)\,.  
\end{equation}
We can use the general result \cref{eq:scalarmodeDF} to compute the energy flux.
\begin{align}
    \dot{E}=\pi(\rho_{E}+p_{E})16\pi^2G^2M^2(2\gamma^2-1)^2 \int_{k}2\pi\delta(k\cdot v)\,\frac{|\omega|}{|\vec k|^2}\delta(\omega^2-c_s^2|\vec k|^2)\,.
\end{align}
The angular integral can be performed trivially, and the delta functions trivialize two other integrals, leaving just a single integral over the magnitude of momentum. Aside from the denominator $|\vec k|^2$, this is precisely the integral one performs to compute the energy flux from Cherenkov radiation, hence the moniker ``acoustic Cherenkov'' radiation used for dynamical friction. Evaluating the integral and reconstructing the force using \cref{eq:forceFromFlux}, we obtain
\begin{equation}\label{eq:fluid_df_result}
    F^{\mu}=-4\pi G^2(\rho_{E}+p_{E})M^2 \left(\frac{\gamma v^{\mu}-u^{\mu}}{\gamma^2-1}\right)\frac{(2\gamma^2-1)^2}{\gamma v}\Theta(v-c_s)\log\left(\frac{k_{max}}{k_{min}}\right)\,,
\end{equation}
which agrees  with the results in \cite{Petrich:1989}.

The ``Coulomb'' log contains both an IR ($k_{min}$) and UV ($k_{max}$) divergence. The IR divergence carries physical information, with a real physical IR cutoff $k_{min}^{-1}\sim L$ corresponding to the cloud size (which is bounded by the Jeans length).  The UV divergence corresponds to a counterterm which we will discuss below. The result \eqref{eq:fluid_df_result} will not hold when $v \sim c_s$, where the particle sits in its own shock wave. In this limit one can't take the infinite time limit first since the argument of the exponent vanishes for collinear radiation. Physically the shock wave formation time becomes infinite. The finite time calculation will have a log divergence in this limit,
since $k \cdot v=|k|\gamma( c_s  - \cos \theta  v)$ in the denominator will have an endpoint singularity in the angular integration. See \cref{eq:viscoussystems} for further discussion of this limit. 

By focusing on the steady-state result, we have taken the total time to be the largest length scale. Corrections to this approximation will scale as $L/(c_sT)$, and for times short compared with the sound crossing time of the cloud the result \cref{eq:fluid_df_result} is not valid. As previously discussed, there is no obstruction to studying the short-time case in this formalism however one will need to include the conservative parts of the force to have a complete result.

 The UV divergence is universal in two ways. Firstly, for the fluid this UV divergence is independent of trajectory. Secondly, as proven in \cref{sec:UVandRG}, the Newtonian limit of \cref{eq:fluid_df_result} holds for any choice of environment. The cutoff will scale with the size of the object, i.e. the scale of the breakdown of the point particle approximation,
$k_{max}=1/R$. \footnote{Recall we are ignoring mass accretion effects,  otherwise the capture scale $\sim GM/v$ would be a relevant short distance scale.}  

\subsection{Fuzzy Dark Matter}

Our proof of the universal UV logarithm relied on an assumption that new length scales in the description of the environment are shorter than the UV cutoff scale. This assumption covers many cases of interest, but not all. One counterexample is fuzzy dark matter, described by a complex scalar field with mass $m_s$ in a homogeneous state at finite number density. This problem has a variety of different scale hierarchies depending on the  de Broglie wavelength $\lambda_{dB} = (m_sv)^{-1}$. For $\lambda_{dB}\ll R$, the wave nature gets effectively averaged out, so the medium can be treated as a collection of classical particles rather than as a coherent wave, however for intermediate wavelengths $R\ll\lambda_{dB}\ll L$,  the wavelike nature of the environment plays an important role.

We take a free scalar field and expand it about the finite density background, $\phi(x) = (\rho_{E}/2m_s^2)^{1/2}e^{-im_s u\cdot x}+\delta\phi(x)$.
This background is a solution to the equation of motion which in the rest frame corresponds to finite chemical potential $m$.

The normalization is chosen so that the leading order stress-energy tensor is in the canonical form
\begin{equation}
    \langle T^{\mu\nu}\rangle = \rho_{E}u^{\mu}u^{\nu}\,.
\end{equation}
The stress-energy correlator is straightforwardly computed in terms of the two-point function of $\phi$. We'll omit the details of the calculation and directly present the energy density spectral density form factor,
\begin{equation}
    \sigma_{\mathcal{E}} = \pi \rho_{E}\left[(\omega+2m_s)^2\textrm{sgn}(\omega+m_s)\delta((\omega+m_s)^2-|\vec k|^2-m_s^2)+(m_s\to -m_s)\right]\,.
\end{equation}
It's straightforward to check that the other form factors satisfy our simple-system criteria \cref{eq:simplecondition} so that the formula for relativistic dynamical friction due to the scalar field is given by $\sigma_{\mathcal{E}}$ alone, via \cref{eq:scalarmodeDF}.

Inserting this form factor into the general expression for dynamical friction, we see that after a change of integration variables the $(m_s\to-m_s)$ term contributes the same as the first term. The energy flux can then be written as 
\begin{equation}
    \dot{E}=16\pi^2 G^2M^2\rho_{E}\frac{(2\gamma^2-1)^2}{\gamma} \int_{k}(2\pi^2)\delta(\omega-\vec{k}\cdot\vec{v})\,\frac{\omega}{|\vec k|^4}
    \,\,(\omega+2m_s)^2\textrm{sgn}(\omega+m_s)\delta((\omega+m_s)^2-|\vec k|^2-m_s^2)\,.
\end{equation}
The quadratic delta function has poles at $\omega=-m_s\pm\omega_{k}$, where $\omega_{k}=\sqrt{m_s^2+|\vec k|^2}$ is the single particle energy. The gapped two-particle pole, $\omega = -m_s-\omega_k$ is kinematically forbidden, since it is not possible to simultaneously satisfy $\omega = \vec{k}\cdot\vec{v} = -m_s-\omega_k$. 

The gapless pole at $\omega=\omega_{k}-m_s$ does contribute, however it is straightforward to see that the two delta functions can only be satisfied for $|\vec k| \leq 2m_s|\vec v|\gamma^2$. Thus there is an implicit UV cutoff set by the inverse de Broglie wavelength. The integral can be straightforwardly evaluated,  giving 
\begin{align}
    \dot{E}=\frac{4 \pi G^2 M^2\rho_{E} (2\gamma^2-1)^2}{\gamma v}\log\left(\frac{2m_sv\gamma }{\Lambda_{IR}}\right)\,,
\end{align}
which agrees precisely with the known result~\cite{Hui:2016ltb}.  

\subsection{Renormalization}\label{sec:renormalization}

The UV divergence in \cref{eq:fluid_df_result} is necessarily absorbed by a counter term in the worldline effective theory. Since this is a dissipative force, it is necessarily a counter term in the CTP worldline EFT which describes fluid-induced friction forces. Details of this EFT on a fixed gravitational background were worked out in \cite{Modrekiladze:2024htc}, and here we will demonstrate how turning on dynamical gravity renormalizes the parameters of that EFT.

At the lowest order in the derivative expansion the only operator on the worldline that is  consistent with the symmetries (see section \ref{WLEFT}) and leads to dissipation is 
\beq
\label{ct}
S_{\text{dis}}
= \int d\lambda \;K(\gamma,\mu)
\frac{u_r \cdot X_a}{\sqrt{\dot{x_r}^2}}\,,
\eeq
with $u_r^{\mu}$ the four-velocity of the fluid and 
\begin{equation}
    \gamma = \frac{u_r \cdot \dot{x}_r}{\sqrt{\dot{x_r}^2}}\,.
\end{equation}
Varying this term in the action produces a force term in the equation of motion
\begin{equation}
    F^{\mu}_{\textrm{dis}} = P^{\mu}_{\,\,\,\nu}u^{\nu}\,K(\gamma,\mu) = (u^{\mu}-\gamma v^{\mu}) K(\gamma,\mu)\,,
\end{equation}
where, as in previous sections, we drop the $r-$subscript after varying the action and fix the gauge $\dot{x}^2=1$. We see that $K(\gamma,\mu)$ describes a friction force on the object, and it is an arbitrary function of the relative velocity of the object and fluid. 

We now have an additional unknown function, $K$, that will in general depend upon the RG scale $\mu$. 
What is interesting is that even if the fluid of interest has vanishing viscosity this Wilson coefficient will be generated by RG running (even from Chandrasekhar's original calculation). That is, as we shall see, the acoustic radiation leads to an effective, scale-dependent, drag coefficient.

To absorb the UV divergence in \cref{eq:fluid_df_result} we must include the $K(\gamma,\mu)$ term in the action with the specific form
\begin{equation}\label{eq:Kbare}
    K_{\textrm{bare}}(\gamma) = \underset{\delta K(\gamma,\mu)}{\underbrace{-4\pi G^{2}(\rho_{E}+p_{E})M^2\frac{(2\gamma^2-1)^2}{(\gamma^2-1)\gamma v}\Theta(v-c_s) \log\left(\frac{k_{max}}{\mu}\right)}} + K_{\mathrm{ren}}(\gamma,\mu)\,.
\end{equation}
With this scheme choice the total force on the object is given by
\begin{equation}\label{eq:totalForce}
    F^{\mu}=-\left(\gamma v^{\mu}-u^{\mu}\right)\left[4\pi G^2(\rho_{E}+p_{E})M^2 \frac{(2\gamma^2-1)^2}{(\gamma^2-1)\gamma v}\Theta(v-c_s)\log\left(\frac{\mu}{k_{min}}\right)+K_{\mathrm{ren}}(\gamma,\mu)\right]\,,
\end{equation}
and the finite remainder $K_{\mathrm{ren}}(\gamma,\mu)$ is determined by matching with the exact UV calculation, e.g. scattering a fluid on a black hole background.

At the order considered here, the renormalized mass may be treated as RG invariant. Requiring the physical force to be independent of $\mu$ then gives a leading-order RG equation for the renormalized Wilson coefficient
\begin{equation}\label{eq:RGequation}
    \frac{dK_{\mathrm{ren}}(\gamma,\mu)}{d\log(\mu)}=-4\pi G^2(\rho_{E}+p_{E})\frac{(2\gamma^2-1)^2}{(\gamma^2-1)\gamma v}\Theta(v-c_s)M^2\,,
\end{equation}
Possible running of $M$ enters at higher orders, particularly in loop diagrams involving finite-size operator insertions, however this is beyond the present calculation. We then see that even in the limit of vanishing viscosity RG running will generate the dissipative Wilson coefficient $K(\gamma,\mu)$ with a leading order beta equation given by the coefficient of Chandrasekhar's Coulomb log (and its relativistic generalization).

\subsection{Matching}

To perform a matching calculation we would need to compare with an exact computation of a compact object, say for a Schwarzschild black hole, moving uniformly through a fluid.  Moreover, to compare precisely, we would need that UV calculation to use the same IR cutoff scheme that we've used, i.e. a hard cutoff in momentum space. To the best of our knowledge, such a result is not available. In lieu of these results, we will demonstrate an example of the matching calculation for a  collisionless dust. 

 The exact results for the scattering of collisionless dust off a black hole are straightforwardly derived by summing the deflection of a bundle of geodesics (see e.g.~\cite{Traykova:2023qyv}). For a collisionless dust the natural cutoff scheme, for both the UV and IR, is in impact parameter space. The geodesics bifurcate into two sets: those which scatter to infinity, and those which plunge into the black hole. 
 The friction force due to the scattering geodesics can be matched to \cref{eq:totalForce} in the dust limit $(p_E,c_s\rightarrow 0)$. The force due to the plunging geodesics must be matched to operators on the worldline describing accretion, and will presented elsewhere~\cite{GMRII}.

 We'll refer to App. B of \cite{Traykova:2023qyv} for details of the exact computation. For our purposes all that is needed is that the force due to scattered geodesics is given by
 \begin{equation}\label{eq:Fexact}
     F_{\mathrm{exact}}^{\mu} = -\left(\frac{\gamma v^{\mu}-u^{\mu}}{\gamma^2-1}\right) 2\pi\rho_{E}(\gamma^2-1)^{3/2}\int_{b_{crit}(v)}^{b_{max}}db\,b[1-\cos(\chi(b))]\,,
 \end{equation}
where $\chi(b)$ is the exact scattering angle of the geodesic. Crucially, there is a physical UV cutoff on the integral given by the scattering-plunge separatrix
\begin{equation}
    b_{crit}(v)=GM \left[\frac{-1+8v^4+\sqrt{1+8v^2}+4v^2(5+2\sqrt{1+8v^2})}{2v^4}\right]^{1/2}=\frac{4GM}{v}\bigg[1+\mathcal{O}(v^2/c^2)\bigg]\,.
\end{equation}
The integral cannot be evaluated analytically, however ref.~\cite{Traykova:2023qyv} provides a numerical fit to the result, 
\begin{equation}\label{eq:traykovaresult}
    F_{\mathrm{exact}}^{\mu}=-4\pi G^2\rho_{E} M^2 \left(\frac{\gamma v^{\mu}-u^{\mu}}{\gamma^2-1}\right)\frac{(2\gamma^2-1)^2}{\gamma v}\left[\log\left(\frac{b_{max}}{R_c}\right)+\mathcal{R}(v)\right]\,,
\end{equation}
where $R_c = \frac{GM}{v^2}$, and the residual function containing general-relativistic corrections is fit by
\begin{equation}
    \mathcal{R}(v) \approx 3.7+9.24v^2-12.16v^3+4.79 v^4-\log\left(\frac{10 (1+v^2)v b_{crit}}{GM}\right)\,,
\end{equation}
with an error less than 1\% for all values of $v$.

To understand the emergence of the length scale $R_c$, it is useful to take the Newtonian limit of \cref{eq:Fexact} and recover Chandrasekhar's result~\cite{Chandrasekhar1943}. In this limit the exact scattering angle is
\begin{equation}
    \chi^{\mathrm{Newt.}}(b) = 2\arctan\left(\frac{GM}{b\,v^2}\right)\,,
\end{equation}
and $R_c$ is recognized as the impact parameter at which the Newtonian scattering is $\pi/2$. Since there is no horizon in the Newtonian limit, $b_{crit}\to 0$, and we can evaluate the impact parameter integral exactly, yeilding
\begin{equation}
    \vec{F}_{\mathrm{exact}}^{\mathrm{Newt.}} = -4\pi G^2 \rho_{E} M^2 \frac{\vec v}{v^3}\log\left(\frac{b_{max}}{R_c}\right)\,,
\end{equation}
in the limit $b_{max}\gg R_c$. So, in the Newtonian case it is $R_c$ which serves as the effective/physical UV cutoff in the RG logarithm. While one might have naively expected that the UV scale in the relativistic computation is given by the black hole accretion scale, $b_{crit}\sim GM/v$, there is actually a much larger Newtonian length scale $R_c \sim b_{crit}\times(c/v)$ which comes in first to stop the running of the UV logarithm.

Given the exact result in \cref{eq:traykovaresult}, we have the requisite UV data to perform a matching calculation for $K(\gamma,\mu)$ to all orders in $G$. To do so, however, also requires us to have an EFT prediction for the dynamical friction force to all orders in $G$. Ostensibly this requires summing infinitely many diagrams, but we will argue that this is not actually the case.

In the case we're presently considering, of a cold dust following geodesic scattering trajectories, to order $G^{n}$ the EFT prediction for the dynamical friction force is 
 \begin{equation}\label{eq:FPM}
     F^{nPM\,\mu} = -\left(\frac{\gamma v^{\mu}-u^{\mu}}{\gamma^2-1}\right) 2\pi\epsilon(\gamma^2-1)^{3/2}\int_{b_{min}}^{b_{max}}db\,b[1-\cos(\chi^{(n-1)}(b))]\,,
 \end{equation}
where the scattering angle is only needed to order $G^{n-1}$,  and after expanding $1-\cos\chi$ only the $G^{n}$ term is retained. Additionally, in the EFT we have an arbitrary UV cutoff $b_{min}$ as the lower bound on the integral, in contrast with the physical scale $b_{crit}$ in \cref{eq:Fexact}. 

For example, inserting the leading (1PM) scattering angle
\begin{equation}
    \chi^{1\,PM}(b)=\frac{2GM}{b} \frac{\left(2 \gamma ^2-1\right)}{\left(\gamma ^2-1\right)}\,,
\end{equation}
\cref{eq:FPM} reproduces our previous leading (2PM) perturbative result \cref{eq:fluid_df_result} exactly, after mapping the cutoff schemes as $k_{max}=b^{-1}_{min}$ and $k_{min} = b_{max}^{-1}$,
\begin{equation}
    F^{2PM\,\mu}=-4\pi G^2\rho_{E}M^2 \left(\frac{\gamma v^{\mu}-u^{\mu}}{\gamma^2-1}\right)\frac{(2\gamma^2-1)^2}{\gamma v}\log\left(\frac{b_{max}}{b_{min}}\right)\,.
\end{equation}
To compute the dynamical friction force at higher orders we would then simply insert the scattering angle, expanded to the desired PM order, and evaluate \cref{eq:FPM}. While this is already a massive simplification over summing diagrams, even this computation is not necessary to perform the matching.\footnote{This is true provided we continue to ignore finite size effects. The inclusion of such effects introduces novel diagrams which are not captured by the probe scattering angle, and which can generate further nontrivial RG flow, e.g. a running of the mass $M$ due to the presence of the fluid.} The reason is that the nPM scattering angle scales as
\begin{equation}
    \chi^{(n)}(b)\sim \left(\frac{GM}{b}\right)^{n}\,,
\end{equation}
so for all subleading PM orders the force is simply proportional to a power law divergence $\left(GM/b_{min}\right)^{n-2}$. 

The complete EFT prediction for the force also involves the worldline operator $K(\gamma,\mu)$,
\begin{equation}
    F_{EFT}^{\mu}=-\left(\gamma v^{\mu}-u^{\mu}\right)\left[4\pi G^2\rho_{E}M^2 \frac{(2\gamma^2-1)^2}{(\gamma^2-1)\gamma v}\log\left(\frac{b_{max}}{b_{min}}\right)+\mathcal{O}\left(GM/b_{min}\right)+K_{\textrm{bare}}(\gamma)\right]\,.
\end{equation}
In \cref{sec:renormalization} we already discussed the counterterm subtraction for the log divergence. Here we can see that since the higher order diagrams are all pure power law divergences they are entirely subtracted away by the counter term for $K$. Had we used dimensional regularization rather than an impact-parameter cutoff scheme the higher-order contributions would have been scaleless integrals and thus identically zero. Thus, to include all higher orders in $G$ we again split the dissipation operator into bare and renormalized parts, as in \cref{eq:Kbare}, but in the counterterm $\delta K$ now also subtracts all of these power law divergences. What is left is a renormalized force
\begin{equation}
    F_{EFT}^{\mu}=-\left(\gamma v^{\mu}-u^{\mu}\right)\left[4\pi G^2\rho_{E}M^2 \frac{(2\gamma^2-1)^2}{(\gamma^2-1)\gamma v}\log\left(\mu b_{max}\right)+K_{\textrm{ren.}}(\gamma,\mu)\right]\,.
\end{equation}
Matching to the exact result, \cref{eq:traykovaresult}, we then obtain the value for the Wilson coefficient at the length scale $R_{c}$
\begin{equation}\label{eq:exactKren}
    K_{\textrm{ren.}}(\gamma,\mu=R_{c}^{-1}) = 4\pi G^2\rho_{E}M^2 \frac{(2\gamma^2-1)^2}{(\gamma^2-1)\gamma v} \mathcal{R}(v)\,,
\end{equation}
and the RG equation \cref{eq:RGequation} (with $p_{E}$ and $c_s$ set to zero) can be used to run to different scales. We'd like to emphasize, this is an exact result for Schwarzschild black holes.

We are now able to address one of our primary goals for this section, applications to non-trivial orbits. At present the exact GR computations of dynamical friction that capture strong-gravity effects on the environment (in e.g. \cite{Vicente:2022ivh, Traykova:2023qyv, Dyson:2024qrq}) are limited to straight-line motion. For phenomenological applications one must lift these results to non-trivial states of motion, however the computation methods in \cite{Vicente:2022ivh, Traykova:2023qyv, Dyson:2024qrq} do not clearly translate. 

Here we have a precise statement of how their results will translate to non-trivial orbits. For an orbit with a characteristic scale $r$ we can use our EFT to systematically compute the dynamical friction force in powers of $GM/r$. At each order there will be UV divergences that must be renormalized. The power law divergences are absorbed into local counterterms, and the logarithmic divergence can be reliably cutoff at a scale $\mu=R_c^{-1}$ provided one also includes the renormalized friction operator, \cref{eq:exactKren}. Doing so, one can trust that they are treating the short-distance physics precisely with no ambiguities around the choice of UV cutoff or the addition of finite counter terms.

\subsection{A comment on viscous systems}\label{eq:viscoussystems}

In first-order (viscous) hydrodynamics the scalar modes are no longer linearly dispersive $\omega\propto k$, but instead they acquire an imaginary diffusive part. Moreover, the vector modes become purely diffusive $\omega \propto -i|\vec k|^2$. Additionally, the spin-2 form factor doesn't have poles. It is instead an analytic function describing a purely viscous response. A natural question is then, how do these viscous effects modify dynamical friction?

To address this we can focus on the scalar modes, and the lessons will generalize to the others. If we include viscous corrections to the hydrodynamic modes, then the energy-density response function becomes
\begin{equation}
   \langle T_{00}T_{00}\rangle^{\textrm{1st order}} = \frac{-(\rho_{E}+p_{E})|\vec{k}|^2}{\omega^2-c_s^2|\vec{k}|^2+i\omega|\vec{k}|^2\gamma_{s}}\,,
\end{equation}
where the sound attenuation constant $\gamma_s$ is proportional to the microscopic relaxation time. 
Taking the imaginary part leads to a Lorentzian weight which scales as $1/k^6$ and suggests that the viscosity effects could soften the UV divergence in the dynamical friction integral.
A careful consideration of the relevant scales and corresponding EFT cutoffs suggests, however, that this softening lies outside of the regime of validity of the EFT.

There is a hierarchy of scales that we must consider. Our point particle EFT has a cutoff length scale (and time scale) $\sim R$ ($\sim R/v$), which are astrophysically large. They are set by the larger of i) the size of the object or ii) the Newtonian gravitational capture radius of the object. The hydrodynamic EFT also has cutoff length/time scale, $l_{\mathrm{mfp}}, \tau_{\mathrm{rel}}$, and these are microscopically small.\footnote{Although we note models like Shakura-Sunyaev, in which case larger scale turbulence is ``integrated out'' to obtain viscous effects on astrophysical scales.} As a consequence of this hierarchy the momentum integral is cutoff at a frequency $\omega_{max}$ such that $\omega_{max}\gamma_{s}\ll1$, and for generic values of $(\omega,|\vec k|)$ the diffusive term is always perturbatively small relative to the dispersive term.

In our integral, we must then expand the integrand in the region for which $(\omega,\vec{k})\sim R^{-1}\ll \gamma_{s}^{-1}$. We then obtain integrals of the form
\begin{equation}\label{eq:hydroregionsexpansion}
    \dot{E}\propto \textrm{Im}\int d^{D}k\,\delta(\omega-\vec{k}\cdot \vec{v})\frac{\omega}{|\vec{k}|^2} \frac{1}{(\omega+i0)^2-c_s^2|\vec{k}|^2}\bigg[1+\sum_{n=1}\left(\frac{-i\omega|\vec{k}|^2 \gamma_s}{(\omega+i0)^2-c_s^2|\vec{k}|^2}\right)^{n}\bigg]\,.
\end{equation}
For all $n\geq1$ the contribution is a scaleless power-law divergence in $D=4$. They are then ``pure counterterm'', i.e. the effects of fluid viscosity on dynamical friction are entirely described by operators in the effective dissipative worldline theory, and there is no long-distance contribution. To be clear, fluid viscosity can indeed contribute to dynamical friction on a compact object but upon performing a matching calculation one will find that the UV computation will be matched perfectly by tree-level calculations for appropriate choices of Wilson coefficients in the effective dissipative worldline theory, without need for further calculations in the IR.  

The above argument holds for generic energy and momenta. However, for speeds very near the speed of sound the leading ``ideal'' term in the hydrodynamic pole can be made arbitrarily small since 
\begin{equation}
    \omega^2-c_s^2|\vec k|^2 = |\vec k|^2 (v^2\cos\theta^2-c_s^2) \ll 1\,.
\end{equation}
Viscous contributions to dynamical friction were analyzed in~\cite{Katz:2019rgf}, where it was argued that for $v\sim c_s$ the first-order hydrodynamic response term grows comparable to the zeroth-order response terms.  Despite this, it is argued that the hydrodynamic expansion is not necessarily breaking down as the Knudsen number remains small so higher order hydrodynamic corrections remain subleading to this viscous term. For nearly sonic velocities then, one cannot perform the expansion in \cref{eq:hydroregionsexpansion} to conclude that viscous corrections are pure counterterm. Indeed,~\cite{Katz:2019rgf} demonstrate that viscous effects in the near-sonic regime smooth out the discontinuous $\theta(v-c_s)$ from the ideal fluid case. The width of this near-sonic region is, however, bounded by
\begin{equation}
    |v^2-c_s^2| \lesssim  \gamma_s/R\,,
\end{equation}
which is incredibly small as a consequence of the hierarchy $l_{\mathrm{mfp}}/R \lll 1$, and likely cannot be resolved.

\section{Conclusions}

In this work we developed a worldline EFT framework for the generalized self-force problem in media and illustrated how it systematically captures both dissipative and conservative effects. We applied the formalism to several environments, including fluids, collisionless matter,
and coherent fields, and found agreement with the literature. Our formalism unifies and generalizes previous calculations in several ways.
\begin{itemize}
\item We give a universal result for the self-force in terms of the retarded  stress energy tensor correlation
function that applies to leading order.
\item We prove  that the famous Chandrasekhar  log is  path independent. Then using a sum rule we prove that  
the log is numerically identical  for all environments, assuming that the UV cut-off for the environment is at a distance scale shorter than the radius of the object. We also show that when this criterion is not met,  as in the case of a coherent field, 
the correct result follows by a  change in the renormalization group scale.
\item  By embedding the EFT in the CTP formalism, it is shown that the UV piece of the aforementioned
Coulomb Log renormalizes (generates) a time reversal violating friction term. The beta function for this
coupling is proportional to the square of the conservative coupling $M$ and will lead to a non-trivial RG flow, assuming there is no non-renormalization theorem for $M$.
\item We have demonstrated how to match exactly the full theoretical result in \cite{Traykova:2023qyv} for the straight-line trajectory,
to extract the value of the frictional coupling $K$, which can then be used to calculate the force for
any choice of trajectory. 
\end{itemize}

We would like to highlight the importance of the WL EFT \cite{GoldbergerRothstein2006} here. In practical applications of dynamical friction calculations to gravitational wave physics, it is nearly always the case that practitioners take the straight-line result, make an educated guess about the value of the Coulomb logarithm, and then insert the force into an orbit evolution code. For rough approximations this is sufficient, but in an era of precision gravitational wave science the uncontrolled errors made in this approach deserve improvement. 

\medskip

\textbf{Note:} As we were about to submit this paper \cite{datta2026} appeared, which discusses related ideas.

\section{Acknowledgments}
B.M.\ and I.Z.R.\ thank Rafael Porto for a helpful discussion. J.W.-G. thanks Soumodeep Mitra for related discussions. 
B.M.\ is supported by the Lightcone Foundation and by the Deutsche
Forschungsgemeinschaft (DFG) under Germany's Excellence Strategy,
EXC 2121 ``Quantum Universe'' (390833306), DESY-26-112. I.Z.R.\ and  J.W.-G.\ are supported by the US Department of Energy grant DE-SC001011.  J.W.-G. is supported by a President’s Postdoctoral Fellowship at CMU.

\appendix

\section{Review of Diffeomorphism-Invariant Formulations of Point-Particle In-In Actions}
\label{app:diff-invariant-in-in}

To manifest diffeomorphism  invariance we begin by assuming that the world-lines are in each other's normal convex neighborhood. That is, for
every pair of points on $x_-$ and $x_+$ there is a unique geodesic between the two. We require that for a given parameterization $x_-(\lambda)$, there exists a family
of geodesics which are indexed by $\lambda$ and parametrized by $s$, $z_\mu(\lambda,s)$, such that $z^\mu(\lambda,s_0)=
x_-^\mu(\lambda)$
and that $z^\mu(\lambda,s_1)$ is a unique point on $x_+$ for all $\lambda$. For a choice of geodesic family
$z_\mu(\lambda,s)$ and a parameterization $x_-(\lambda)$ we then have an induced parameterization $x_+(\lambda)$. The
two-fold reparameterization invariance of the dynamics will manifest itself in a reparameterization
invariance of $z_\mu$ in addition to invariance under the choice of geodesic family. For a given point
$x_-(\lambda)$ we could have considered any geodesic $z_\mu(\lambda,s)$ which connects to a point on $x_+$.

We then introduce Synge's world function
\begin{equation}
\label{sigma}
\sigma\!\bigl(x_-(\lambda),x_+(\lambda)\bigr)
=\frac{(s_1-s_0)}{2}\int_{s_0}^{s_1} ds\;
g_{\mu\nu}\!\bigl(z(\lambda,s)\bigr)\,
\frac{dz^\mu(\lambda,s)}{ds}\,
\frac{dz^\nu(\lambda,s)}{ds},
\end{equation}
which is invariant under affine re-parameterizations of $s$, and is equal to one-half the squared geodesic
distance between $x_-(\lambda)$ and the corresponding point $x_+(\lambda)$. The quantity
\begin{equation}
\epsilon \equiv -g_{\mu\nu}\!\bigl(z(\lambda,s)\bigr)\,
\frac{dz^\mu(\lambda,s)}{ds}\,
\frac{dz^\nu(\lambda,s)}{ds}
\end{equation}
is a constant since we have chosen  a parameterization of the geodesic  which is affine to the proper time, so the norm of the tangent is preserved , and therefore numerically 
\begin{equation}
\sigma\!\bigl(x_-(\lambda),x_+(\lambda)\bigr)
=-\frac{1}{2}(s_1-s_0)^2\,\epsilon.
\end{equation}

Derivatives of this function with respect to $x_-$ are tangent to the geodesics,
\begin{equation}
\label{xa}
\frac{\partial}{\partial x_-^\mu} \sigma(x_-,x_+)
=-(s_1-s_0)\,g_{\mu\nu}\!\bigl(x_-(\lambda)\bigr)\,
\left.\frac{dz^\nu(\lambda,s)}{ds}\right|_{s=s_0},
\end{equation}
is a (co)vector on the tangent space at $x_-$, and $z(s_0)=x_-$.

An important property, which we will later use, is that this vector squares to the world function:
\begin{equation}
g^{\mu\nu}(x_-)\,\partial_\mu\sigma(x_-,x_+)\,\partial_\nu\sigma(x_-,x_+)
=2\,\sigma(x_-,x_+).
\end{equation}

This construction allows us to introduce Keldysh variables. Let us define $x_r^\mu$ to be the point
along the curve $z^\mu(\lambda,s)$ which, for each $\lambda$, is one-half the geodesic distance from
$x_-$ to $x_+$. i.e. if $s$ is taken as the proper length parameter (for space-like geodesics),
\begin{equation}
x_r^\mu(\lambda)= z^\mu\!\left(\lambda,\frac{s_1+s_0}{2}\right).
\end{equation}
Then define
the Keldysh $x_a$ variable to be
\begin{equation}
x_{a\,\mu} \equiv -2 \frac{\partial}{\partial x_{r}^\mu}\,\sigma(x_r,x_+).
\end{equation}
So $x_a$ lives at the point $x_r$.
Whereas $x_r^\mu$ is not a vector, just a coordinate, we see that $x_{a\mu}$ is indeed a (co)vector. Using the
definition of the world function it is easy to check that
\begin{equation}
g_{\mu\nu}\,x_a^\mu x_a^\nu
= 8\,\sigma(x_r,x_+)
= 2\,\sigma(x_-,x_+),
\end{equation}
which is precisely the squared geodesic distance between $x_-$ and $x_+$, in agreement with the
standard flat spacetime definition.

To write down the action in terms of these curved space generalizations of the Keldysh variables we first need to define the two reparameterization invariances on our new variables.
In the original variables we had RPI under changes of affine parameters $\lambda_\pm$. Infinitesimally the diagonal (D) and off-diagonal (OD) group actions 
correspond to $\delta_D \lambda: \delta \lambda_-
=\delta \lambda_+=\epsilon$ and $\delta_{OD}\lambda: \delta \lambda_+=\epsilon=-\delta \lambda_-$. The diagonal group action corresponds to the reparameterization of $x_r$, while
the off diagonal action changes both points at the end of the geodesic while holding $x_r$ fixed and changes the geodesic whose mid point is $x_r(\lambda)$.  Thus $x^\mu_a,x^\mu_r$ are scalar fields under the diagonal group, i.e. this action does not change the curve. Under the off-diagonal RPI $x_r$ is unchanged but since the
geodesics change $x^\mu_a$ is not a scalar. However, the variation of $x_a$ under the off-diagonal RPI 
\beq \delta_{OD} x_a^{\mu} \propto \dot{x}_r^\mu 
\eeq
This is most easily seen by going to flat space and performing an infintesimal variation $\delta_{OD} (x_+-x_-) \sim (\dot x_++\dot x_-)$.
Thus, we can form a scalar by projecting $x_a$ onto the direction orthogonal to $\dot x_r^\mu$.

Now when we vary the action we will do so with respect to the flat space $x_a^f$,
so it behooves us to express 
$x_a$ in terms of $x_a^f$, which we can order in powers of $G$. We will set $x_a$ to live at the midpoint, so that $z(-1/2)=x_-$ and $z(1/2)=x_+$. Then expanding around the mid-point write
\beq
\label{mid}
z^\mu(s)= x_r^\mu+ A^\mu s+\frac{1}{2} s^2 B^{\mu}+\mathcal{O}(s^3).
\eeq
Then using the geodesic equation we have up to order $s^2$ 
\beq
B^\mu= - \Gamma^\mu_{\alpha \beta}(x_r) A^\alpha A^\beta.
\eeq
Then using  the mid point defintion of $x_a$, (\ref{mid})
\beq
A^\mu=x_a^\mu\,,
\eeq
such that,
\bea
 x_+&=&x_r^\mu+\frac{1}{2}x_a^\mu - \frac{1}{8}\Gamma^\mu_{\alpha \beta}(x_r) x_a^\alpha x_a^\beta \nn \\
 x_-&=& x_r^{\mu}-\frac{1}{2}x_a^\mu - \frac{1}{8}\Gamma^\mu_{\alpha \beta}(x_r) x_a^\alpha x_a^\beta,
\eea
So we see that 
\beq
x_a^\mu
=
x_{a,f}^\mu
+
O\!\left(x_{a,f}^3\right),
\eeq
so that as long as we are not interested in stochastic noise, we may simply treat $x_a^\mu$ as a vector. Notice that had we not expanded around the central point the $x_a^2$ terms would not have canceled.
Thus the mid-point definition is a more convenient choice of coordinates.

\bibliography{references} 
\end{document}